\def\year{2023}\relax

\documentclass[letterpaper]{article} 
\usepackage{aaai23}  
\usepackage{times}  
\usepackage{helvet} 
\usepackage{courier}  
\usepackage[hyphens]{url}  
\usepackage{graphicx} 
\usepackage{natbib}  
\usepackage{caption} 
\usepackage{enumitem}
\usepackage{nameref}

\usepackage{booktabs} 
\usepackage{multirow}
\usepackage{makecell}

\usepackage{csquotes}
\usepackage{amsmath}
\usepackage{bm}
\usepackage{outlines}

\usepackage{etoolbox}
\robustify\bfseries

\usepackage{siunitx}
\usepackage{subcaption}

\usepackage{graphicx}

\usepackage{placeins}

\usepackage[dvipsnames]{xcolor}

\usepackage{pifont}
\usepackage{array}
\newcolumntype{P}[1]{>{\centering\arraybackslash}m{#1}}
\newcolumntype{R}[1]{>{\raggedright\arraybackslash}m{#1}}
\newcolumntype{L}[1]{>{\raggedleft\arraybackslash}m{#1}}

\usepackage{xspace}
\newcommand\ie{i.\,e.\xspace}
\newcommand\eg{e.\,g.\xspace}

\usepackage{url}

\makeatletter
\renewcommand{\fps@figure}{htb}         
\renewcommand{\fps@table}{htb}         
\makeatother 

\usepackage[ruled]{algorithm2e}
\title{Finding Qs: Profiling QAnon Supporters on Parler}

\author{Dominik Bär\textsuperscript{\rm 1,2}, 
        Nicolas Pröllochs\textsuperscript{\rm 3},
        Stefan Feuerriegel\textsuperscript{\rm 1,2} \\}
\affiliations{
    \textsuperscript{\rm 1}LMU Munich, Munich, Germany \\
    \textsuperscript{\rm 2}Munich Center for Machine Learning, Munich, Germany \\
    \textsuperscript{\rm 3}JLU Giessen, Giessen, Germany \\
    baer@lmu.de, nicolas.proellochs@wi.jlug.de, feuerriegel@lmu.de
}

\nocopyright

\begin{document}

\maketitle

\begin{abstract}
The social media platform ``Parler'' has emerged into a prominent fringe community where a significant part of the user base are self-reported supporters of QAnon, a far-right conspiracy theory alleging that a cabal of elites controls global politics. QAnon is considered to have had an influential role in the public discourse during the 2020~U.S.~presidential election. However, little is known about  QAnon supporters on Parler and what sets them aside from other users. Building up on social identity theory, we aim to profile the characteristics of QAnon supporters on Parler. We analyze a large-scale dataset with more than 600,000 profiles of English-speaking users on Parler. Based on users' profiles, posts, and comments, we then extract a comprehensive set of user features, linguistic features, network features, and content features. This allows us to perform user profiling and understand to what extent these features discriminate between QAnon and non-QAnon supporters on Parler. Our analysis is three-fold: (1)~We quantify the number of QAnon supporters on Parler, finding that 34,913 users {(\SI{5.5}{\percent} of all users)} openly report supporting the conspiracy. (2)~We examine differences between QAnon vs. non-QAnon supporters. We find that QAnon supporters differ statistically significantly from non-QAnon supporters across multiple dimensions. For example, they have, on average, a larger number of followers, followees, and posts, and thus have a large impact on the Parler network. (3)~We use machine learning to identify which user characteristics discriminate QAnon from non-QAnon supporters. We find that user features, linguistic features, network features, and content features, can -- to a large extent -- discriminate QAnon vs. non-QAnon supporters on Parler. In particular, we find that user features are highly discriminatory, followed by content features and linguistic features.
\end{abstract}

\section{Introduction}

The social media platform ``Parler'' has emerged into a prominent fringe community, where a significant user base identifies with far-right viewpoints. Parler was founded in August~2018 and promotes open self-expression and free speech \cite{Aliapoulios.2021, Otala.2021}. Its user base grew during the 2020 U.S. presidential election and, in January 2021, counted 13.255\,M users \cite{Aliapoulios.2021}.  

In recent years, Parler gained widespread public interest when conservative thought leaders endorsed the platform as an alternative to mainstream social media platforms \cite{Aliapoulios.2021}. As such, a large portion of the content on Parler is characterized by conservative viewpoints. For example, many discussions revolve around former U.S. President Donald Trump, the 2020 U.S. presidential election, and religion \cite{Aliapoulios.2021}. In addition, Parler users were associated with the storming of the U.S. Capitol on January~6,~2021 \cite{Hitkul.2021}. The latter eventually led to the removal of the Parler app from both the Google Play Store and Apple's App Store, as well as Amazon stopping to host the website \cite{Aliapoulios.2022}. As a result, Parler went offline on January~11,~2021 but was reinstated on February~15,~2021 after finding a new hosting service \cite{Robertson.2021}.

Besides conservative viewpoints, Parler users frequently identify with various conspiracy theories \cite{Aliapoulios.2022, Hitkul.2021}. Different from (necessary) healthy skepticism of government and elites, beliefs in false narratives and conspiracies can pose serious threats to democracy and public safety \cite{Geissler.2022, Sternisko.2020}. As a particularly concerning example, Parler users have been associated with QAnon, a far-right conspiracy theory in which supporters allege that a cabal of elites controls global politics, conspires in a pedophile ring, and engages in destroying society \cite{Aliapoulios.2022, Amarasingam.2020, Hanley.2022}. QAnon combines several conspiracy theories, and, as such, research has characterized it as a ``meta narrative'' \cite{Zuckerman.2019} or ``super-conspiracy theory'' \cite{Papasavva.2021} that quickly gained a large base of supporters and expanded globally \cite{Hoseini.2021}.

The prominence of QAnon has led to widespread concerns. One reason is that QAnon supporters were frequently associated with violent incidents \cite{Hoseini.2021}. For example, several QAnon supporters participated in the storming of the U.S. Capitol on January~6, 2021 \cite{Aliapoulios.2022}. As a result, the U.S. Federal Bureau of Investigation (FBI) has labeled QAnon a national security threat \cite{Amarasingam.2020}. Moreover, major social media networks such as Reddit, Twitter, and Facebook have banned QAnon-related content from their platforms \cite{Papasavva.2021}, and, as a consequence, QAnon supporters migrated -- to a large extent -- to fringe communities such as Parler \cite{Aliapoulios.2022}. However, a user-centric view profiling QAnon supporters on Parler is missing and presents our novelty.

Despite the above, the total size of the QAnon community on Parler remains unclear. Yet, understanding the size of the community is crucial to evaluate the security threat that emanates from QAnon on Parler and motivates us to quantify the number of self-reported QAnon supporters on the platform. Furthermore, little is known about characteristics that discriminate QAnon supporters from non-supporters. Parler is characterized by a largely right-leaning community \cite{Otala.2021}, which leads to a rather homogeneous user base compared to mainstream platforms such as Twitter or Facebook. As a consequence, QAnon supporters and other users on Parler might -- at least to some extent -- share a similar social identity and might hardly be distinguishable. Social identity theory argues that the behavior of people depends on -- and varies across -- group membership \cite{Tajfel.1986}, which can strengthen beliefs in conspiracy theories and drive behavior \cite{Sternisko.2020}. Adherents of conspiracy theories might, for example, want to improve their self-image or the reputation of the group they belong to and find explanations for their beliefs \cite{Sternisko.2020}. In fact, previous research on Facebook and Twitter showed that conspiracy theorists behave differently compared to other users \cite{Bessi.2015, Sharma.2022}. However, it is unknown if the same holds for QAnon users on Parler. Gaining such knowledge would have important practical implications as it might help surveillance of conspiracy theorists and improve early detection of upcoming security threats.

\textbf{Research Questions:} In this paper, we profile self-reported QAnon supporters on Parler based on a large-scale public snapshot of the Parler network \cite{Aliapoulios.2021}. In particular, we seek to answer the following research questions (RQs):
\begin{itemize}
	\item \textbf{RQ1:} \emph{How many users on Parler are self-reported QAnon supporters?}
	\item \textbf{RQ2:} \emph{How do user characteristics differ between self-reported QAnon supporters and other users on Parler?}
	\item \textbf{RQ3:} \emph{Which user characteristics allow machine learning to discriminate self-reported QAnon supporters from other users on Parler?}
\end{itemize}

\noindent
\textbf{Contributions:} By addressing the above RQs, we aim at profiling the characteristics of self-reported QAnon supporters on Parler. Our contribution is three-fold: (1)~We quantify the number of self-reported QAnon supporters on Parler. (2)~We make use of a user's profile, friendships, posts, and comments, based on which we extract a comprehensive set of user features, linguistic features, network features, and content features. We then compare QAnon vs. non-QAnon supporters along these features to identify user characteristics that distinguish both groups. (3)~We use machine learning methods to identify which user characteristics can be leveraged to discriminate QAnon supporters from non-QAnon supporters.

\section{Related Work}

\textbf{User Profiling:} Social networks attract a large and diverse user base. While users disclose some personal information, other characteristics can be kept private. As a result, research has leveraged data from social media platforms to profile different user groups. For example, one study seeks to understand what distinguishes verified from non-verified users on Twitter \cite{Paul.2019}. Other works use public information of users to predict their gender, age, and geographic origin, even if such information is kept private \cite{Burger.2011, Rao.2010}. Even other works leverage personal information to learn about the socioeconomic status and income of users \cite{Lampos.2016, PreotiucPietro.2015, Rao.2010}. Previous literature has also studied the characteristics of users engaging in online abuse \cite{Chatzakou.2017b, Chatzakou.2017, ElSherief.2018, Gorrell.2018, Hua.2020, Maity.2018, Ribeiro.2018}. Here, we add by profiling QAnon supporters. 

\vspace{0.2cm}
\noindent
\textbf{Fringe Online Communities:} Fringe online communities (sometimes also referred to as alt-tech) are characterized by open self-expression and free speech \cite{Aliapoulios.2021, Bar.2023, Papasavva.2021}. Examples of corresponding social media platforms are Gab, 4chan, Voat, BitChute, Gettr, and Parler. Over the last years, they attracted large numbers of new users that were dissatisfied with or banned from mainstream communities such as Twitter or Facebook (\ie, de-platforming) as many of these are increasingly invested in content moderation efforts \cite{Aliapoulios.2021, Otala.2021}. As such, fringe communities attract a different user base compared to mainstream communities characterized by extreme viewpoints. In particular, many fringe communities host users from the far-right political spectrum or conspiracy theorists \cite{Aliapoulios.2021, Aliapoulios.2022, Papasavva.2021, Zannettou.2018}. 

There is a growing interest in computational social science to research the behavior of fringe communities at Gab \cite{Ali.2021, Zannettou.2018}, 4chan \cite{Hine.2017, Papasavva.2020}, Voat \cite{Papasavva.2021}, BitChute \cite{Trujillo.2020}, Gettr \cite{Paudel.2022}, or Parler \cite{Aliapoulios.2021, Baines.2021, Jakubik.2023, Munn.2021, Otala.2021, Pieroni.2021}. In this work, we focus on Parler due to its alleged role in inciting violence, disseminating extreme far-right content \cite{Aliapoulios.2021, Hitkul.2021}, and its relevance for hosting conspiracy theories as is the case for QAnon \cite{Aliapoulios.2021}.

\vspace{0.2cm}
\noindent
\textbf{Parler:} Parler operates as a microblogging service similar to Twitter and has grown a significant fringe online community over the last years. The Parler community is characterized by extreme viewpoints and accused of partially coordinating the storming of the U.S. Capitol on January~6, 2021 \cite{Aliapoulios.2021}. Parler is subject to heavy political polarization \cite{Munn.2021, Otala.2021}. In fact, previous research has shown that many politicians from the Republican Party, as well as their followers, migrated from Twitter to Parler during the 2020 U.S. presidential election and around the storming of the U.S. Capitol \cite{Otala.2021} dissatisfied with increasing content moderation on Twitter. The authors suggest that this migration has contributed to a Parler community that is -- to a large extent -- right-leaning.

Research examining the social media platform Parler has only recently received traction. One stream of literature is interested in cross-platform comparisons, whereby content from Parler is compared against, \eg, Twitter. Here, findings suggest that Parler encompasses views and emotions that differ from those shared on Twitter, especially with regards to the storming of the U.S. Capitol on January~6, 2021 \cite{Hitkul.2021,Jakubik.2023}. For example, many Parler users are supportive of the strongly conservative ideology of the rioters, whereas most users on Twitter condemned the riots. 

Another stream of literature has studied \emph{what} content is posted on Parler. On the one hand, content appears to frequently reference former U.S. President Donald Trump \cite{Aliapoulios.2021}. On the other hand, content on Parler was found to cover various conspiracy theories \cite{Aliapoulios.2021, Pieroni.2021}. In this regard, content on Parler was also found to make frequent references to QAnon \cite{Aliapoulios.2021, Sipka.2022}. However, other than that, there is a paucity of works analyzing the online behavior of QAnon supporters on Parler.

\vspace{0.2cm}
\noindent
\textbf{QAnon:} QAnon emerged supposedly in 2017 \cite{Papasavva.2021}, when an anonymous user named ``Q'' posted a thread named ``Calm before the storm'' stating that former U.S. President Donald Trump is leading the fight against a cabal of elites \cite{Aliapoulios.2022}. Ever since, cryptic pieces of information have appeared online, which are decrypted by QAnon supporters informing them about the fight against the cabal of elites \cite{Aliapoulios.2022}. 

A large part of the literature on QAnon focuses on the online diffusion of the conspiracy theory \cite{Aliapoulios.2022, Hanley.2022, Hoseini.2021, Priniski.2021}. Scholars have analyzed posts attributed to ``Q'' (so-called ``Q drops'') from different aggregation sites \cite{Aliapoulios.2022}. However, they find only little agreement between the Q drops from different websites, thus implying that Q drops are written by multiple people. In this regard, prior literature has provided a detailed overview of QAnon discussions on, \eg, Voat \cite{Papasavva.2021}, or compared content across platforms \cite{Sipka.2022}. Furthermore, there exists evidence that the QAnon conspiracy theory is also discussed on other platforms such as Twitter \cite{Sharma.2022}, YouTube \cite{Miller.2021}, 8kun \cite{Aliapoulios.2022}, or Parler \cite{Aliapoulios.2022}.

Previous research has also studied QAnon supporters \cite{Engel.2022, Papasavva.2021, Priniski.2021, Sharma.2022}. In the case of Twitter, QAnon supporters engage in active discussions and effectively circumvent content moderation \cite{Sharma.2022}. Moreover, QAnon supporters tend to disseminate rather than produce content \cite{Priniski.2021}. Similarly, QAnon supporters actively participated in discussions on Reddit and frequently shared low-quality links \cite{Engel.2022}. For the QAnon-related subverse on Voat, it was found that only a few users are responsible for writing the majority of posts, while comments are made by a large base of subscribers \cite{Papasavva.2021}. However, none of these studies profile QAnon supporters on Parler, especially not with the objective of understanding which user characteristics discriminate QAnon supporters from non-QAnon supporters.

\vspace{0.2cm}
\noindent
\textbf{Research Gap:} Little is known about which user characteristics can discriminate QAnon supporters from non-QAnon supporters. In this paper, we close this research gap by profiling the characteristics of QAnon supporters on Parler. 

\section{Data}

Our analysis is based on a large-scale public snapshot of the Parler social network \cite{Aliapoulios.2021}. Specifically, we collect data from 638,865 English-speaking users between the founding of Parler in August~2018 to its shutdown on January~11, 2021. These users posted approximately 158\,M posts and 42\,M comments. Overall, the size of our dataset is comparatively large-scale, especially when compared with user profiling in other contexts \cite[\eg,][]{Lampos.2016, Matero.2019, PreotiucPietro.2015}.

Our dataset contains information on a user's friendship network (\ie, followers, followees), online activity (\ie, number of posts, number of comments), and popularity (\ie, number of upvotes per post, impressions per post, up/downvotes per comment). Furthermore, for each account, we collect information about the date on which users joined Parler and the user bios. User bios on Parler are short self-descriptions of Parler users. They appear on the front page of a user's profile and are similar to user bios on Twitter. For the purpose of our study, we only focus on users with non-empty bios. Our data also contains the content posted by each user from August~2018 to January~11, 2021. 

\section{Methods\footnote{Code is available via \url{https://github.com/DominikBaer95/Parler_QAnon_UserProfiling}.}}
\label{sec:methods}

\subsection{Theoretical Motivation}

In this work, we aim to quantify the number of self-reported QAnon supporters on Parler (\textbf{RQ1}); compare them to non-QAnon supporters across different user characteristics (\textbf{RQ2}); and, eventually, examine which user characteristics can discriminate both QAnon vs. non-QAnon supporters using machine learning (\textbf{RQ3}). 

There are several reasons why we may or may not find differences with regard to \textbf{RQ2}/\textbf{RQ3}. On the one hand, the political orientation on Parler is largely right-leaning \cite{Otala.2021}. This leads to a generally more homogeneous user base compared to mainstream communities such as Twitter or Facebook (where, in the latter, users are still exposed to diverse ideological content \cite{Bakshy.2015}). Within Parler's right-leaning environment, users largely share similar views on many matters, such as their support for Donald Trump. Hence, QAnon supporters on Parler might exhibit -- at least to some extent -- the social identity of non-supporters. This would imply that QAnon supporters and other users are non-distinguishable; we would thus see no differences with regard to \textbf{RQ2} and \textbf{RQ3}. On the other hand, social identity theory argues that the behavior of people varies across different groups \cite{Tajfel.1986}. Hence, people are drawn towards conspiracy theories due to specific social identity motivators \cite{Sternisko.2020}. Here, adherents might want to improve their self-image or the reputation of the group they belong to and find explanations for their environment \cite{Sternisko.2020}. Hence, adherents of conspiracy theories might behave differently to align with these motivations. In fact, previous research on Facebook showed that conspiracy theorists behave differently (\eg, share more content related to their views) as compared to other communities \cite{Bessi.2015}. In addition, research from Twitter points out that QAnon supporters are particularly active in engaging on the platform \cite{Sharma.2022}. Overall, this would imply that there are indeed differences with regard to \textbf{RQ2} and \textbf{RQ3}. Resolving this opposition motivates our work to profile user characteristics of QAnon supporters on Parler.

\subsection{Identifying QAnon Supporters}
\label{sec:user_identification}

To quantify the size of the QAnon community on Parler, we first classify users into whether they self-report as QAnon supporters. For this, we follow the approach in \citet{Sharma.2022} and apply a keyword list to a user's bio in order to classify users. The theoretical reasoning behind this approach is that the user bios are supposed to reflect the personal and social identity of a user \cite{Li.2020, Pathak.2021, Rogers.2021}. As such, user bios are predictive of one's identity, including gender \cite{Burger.2011, Pathak.2021}, personal interests \cite{Ding.2014}, and political orientation \cite{Pathak.2021, Rogers.2021}, while we here adapt the approach to supporters of conspiracy theories.

We adopt an extensive list of QAnon-specific keywords from prior literature \cite{Sharma.2022} to identify QAnon supporters. These capture different aspects of the underlying conspiracy theory, including the final fight against the cabal (\eg, ``thestorm'', ``thegreatawakening''), the alleged child sex trafficking ring run by the cabal (\eg, ``saveourchildren'', ``adrenochrome''), or related conspiracy theories (\eg, ``pizzagate'', ``obamagate''). Hence, a user is counted as a \emph{QAnon supporter} if one of the keywords appears on a user's bio. Otherwise, we refer to the user as \emph{non-QAnon supporter}. Example user bios of QAnon vs. non-QAnon supporters are shown in Tbl.~\ref{tbl:examples_bio}. 

\vspace{0.2cm}
\noindent
\underline{\textbf{Validation Study:}} We validate the reliability of the above keyword approach as follows. Specifically, we performed a user study on the online survey platform Prolific (\url{www.prolific.co})  with $n=7$ participants in order to assess whether our keyword approach leads to similar results as human assessment. Prior to our user study, all participants were trained in our task by providing them with background information on the QAnon conspiracy theory (and what sets it apart from conservative ideology). Subsequently, the participants were shown 100 randomly sampled user bios (50 QAnon and 50 non-QAnon) and then asked to assess to what extent the corresponding identifies with QAnon. For this, we use a Likert scale ranging from $-3$ to $+3$, where $-3$ corresponds to a user who does not identify with QAnon at all, while $+3$ refers to a strong identification. 

We find a statistically significant interrater agreement in terms of Kendall's $W$ ($W=0.76$; $p<0.01$). We then evaluate to what extent the rater assessments and our keyword approach agree. For this, we first map the Likert scale onto a binary label (QAnon yes/no), whereby a user bio is classified as QAnon (non-QAnon) if a rater has labeled the user bio with a Likert rating of $+1$ or higher (zero or below). We find a large overlap between our keyword approach and the human assessment. In fact, we find that only 4\,\% of QAnon and 2\,\% of non-QAnon supporters were classified incorrectly by the keyword approach, whereas the majority of class labels are in agreement\footnote{We find that misclassified accounts are related to other conspiracy theories, yet, with no other reference to QAnon. For example, users reference the keyword ``plandemic'' which might not provide enough evidence for human annotators to classify the profile as QAnon. To quantify how this would extrapolate to the whole sample, we checked for the number of profiles only mentioning the keyword ``plandemic'' with no other reference to QAnon in their bios: This only accounts for 99 of all accounts in our sample. This thus corroborates the reliability of our approach.}. Consequently, our keyword approach can reliably discern (non-)QAnon supporters on Parler.

\begin{table}
	\small
	\linespread{0.5}
	\begin{tabular}{p{0.18\linewidth}p{0.73\linewidth}}
		\toprule
		\textbf{Class label} & \textbf{Bios (examples)}\\
		\midrule
		\multirow{3}{*}{\vspace{-5em} QAnon} & \emph{``Truth Seeking Psychic Passionate About Revealing False Narratives Perpetrated by The 1\,\% \#WWG1WGA Here To Help Provide Insight And Awaken Humanity \#The Great Awakening''}\\
		\cmidrule(lr){2-2}
		& \emph{``Lover of God, Family, Country, and First Responders. \#TrumpTrain \#WWG1WGA''} \\
		\cmidrule(lr){2-2}
		& \emph{``Conservative. Trump supporter. Q army''}\\
		\midrule
		\multirow{3}{*}{\vspace{-4em} Non-QAnon} & \emph{``a fair-minded thinker that believes in one constitution and equal justice for all.''} \\
		\cmidrule(lr){2-2}
		& \emph{``Believer, Patriot, Conservative, Father, Friend''} \\
		\cmidrule(lr){2-2}
		& \emph{``I am a patriot, an American. I support our president. I will never kneel before our flag...''} \\
		\bottomrule
	\end{tabular}
	\caption{Examples of user bios classified as QAnon and non-QAnon supporters. Class labels via keyword matching.}
	\label{tbl:examples_bio}
\end{table}

\subsection{Feature Extraction}
\label{sec:feature_extraction}

We extract a comprehensive set of features that should characterize our user base (see Tbl.~\ref{tbl:features}), namely (1)~user features, (2)~linguistic features, (3)~network features, and (4)~content features as follows.

\vspace{0.2cm}
\noindent
\textbf{(1)~User Features:} To characterize users, we rely on information on a user's friendship network. This information was discriminatory in other research inferring verified status \cite{Paul.2019}, the socioeconomic circumstances \cite{Lampos.2016}, or discriminating bots and human users \cite{Kudugunta.2018} on Twitter. Consistent with \cite{Kudugunta.2018, Lampos.2016, Paul.2019}, we also include information on a user's activity on Parler, namely, the number of posts and comments as well as the account age.

\vspace{0.2cm}
\noindent
\textbf{(2)~Linguistic Features:} We extract linguistic characteristics that capture the style (\ie, the \emph{how}) with which content is written. Previously, such linguistic features were found to be a reliable predictor of group membership in other settings \cite{Hu.2016, Khalid.2020}. Analogous to prior literature \cite{Chatzakou.2017, Khalid.2020, Lampos.2016, Paul.2019}, we thus control for handles, hashtags, external URLs, long words, and part-of-speech tags (POS). 

We also compute the sentiment and the levels of toxicity, identity attacks, insult, profanity, and threat of content on Parler. Here, we follow prior research \cite{Aliapoulios.2022, Naumzik.2022, Papasavva.2021, Prollochs.2021, Prollochs.2021b, Prollochs.2023} and extract sentiment using the NRC dictionary \cite{Mohammad.2021} and use Google's Perspective API \cite{Jigsaw.2022} to extract the other features (\ie, toxicity, identity attacks, etc.). Overall, this is motivated by the fact that sentiment was found to be discriminatory in similar applications and that different levels of toxicity and threat for QAnon users compared to non-QAnon supporters were observed on other platforms \cite{Aliapoulios.2022, Papasavva.2021, Paul.2019} 
	
We further calculate the average stance of a user towards QAnon. This accounts for the fact that sentiment and stance are not necessarily correlated \cite{ALDayel.2021}. Inspired by prior research \cite{Kawintiranon.2021}, we pre-trained BERT on a large corpus of $\sim$5\,M posts from Parler and added the most important stance tokens towards/against QAnon to the original BERT vocabulary. Overall, this should allow our model to capture Parler-specific language more accurately compared to using standard BERT for the downstream task of stance detection \cite{Kawintiranon.2021}. Subsequently, we fine-tuned the new language model on a sample of 1250 stance-labeled posts and computed the average stance of a user towards QAnon (details are in our GitHub). 

Lastly, we include features extracted using the LIWC 2015 \cite{Pennebaker.2015} and the Empath library \cite{Fast.2016}, which provide a multifaceted view on various lexical categories (\eg, negative/positive emotions, aggression) and were used in various applications studying online content \cite{Kratzwald.2018, Maarouf.2022, Ribeiro.2018, Robertson.2022}.

\vspace{0.2cm}
\noindent
\textbf{(3)~Network Features:} We extract network features as a representation of how users are connected on Parler. Following prior research characterizing hateful users on Twitter \cite{Chatzakou.2017,Ribeiro.2018}, we compute betweenness, eigenvector, and in-/out-degree centrality for each user in the repost network.

\vspace{0.2cm}
\noindent
\textbf{(4)~Content Features:} For each post and comment, we extract text embeddings as a representation of the textual content. In doing so, each post and comment is cleaned and tokenized; \ie, we replace contractions and remove hashtags, handles, emojis, and other alphanumeric characters. Next, we extract text embeddings from the standard SBERT model \emph{all-MiniLM-L6-v2} with mean pooling \cite{Reimers.2019}. In particular, we map each post and comment on a 383-dimensional vector. SBERT is based on Google's BERT model \cite{Devlin.2019} and is specifically trained to create meaningful embeddings for short text and chosen because it is computationally more efficient and showed significantly better results on common benchmarks \cite{Reimers.2019}. Subsequently, we follow prior literature \cite{Yu.2016} and average all sentence embeddings corresponding to a specific user. This provides a representation of the textual content posted by each user.

\begin{table*}[ht!]
	\small
	\setlength{\tabcolsep}{2pt}
	\begin{tabular}{lll}
		\toprule
		\textbf{Feature Group} & \textbf{Dimension} & \textbf{Description} \\
		\midrule
		\multirowcell{9}[0pt][l]{User \\ features} & \emph{Followers} & Number of followers per user\\
		& \emph{Followees} & Number of followees per user \\
		& \emph{Posts} & Number of posts per user \\
		& \emph{Comments} & Number of comments per user \\
		& \emph{Impressions (post)} & Average number of impressions per post \\
		& \emph{Upvotes (post)} & Average number of upvotes per post \\
		& \emph{Upvotes (comment)} & Average number of upvotes per comment \\
		& \emph{Downvotes (comment)} & Average number of downvotes per comment \\
		& \emph{Account age} &  Time passed since a user joined Parler \\
		\midrule
		\multirowcell{16}[0pt][l]{Linguistic \\ Features} & \emph{Handles} & Frequency of handles per post/comment \\
		& \emph{Hashtags} & Frequency of hashtags per post/comment \\
		& \emph{URLs} & Frequency of URLs per post/comment \\
		& \emph{Long words} & Frequency of long words ($\geq$ 6 letters) per post/comment \\
		& \emph{POS} & Frequency of tokens per post/comment identified as part of speech after POS tagging$^{1}$\\
		& \emph{Stance} & Average stance of user towards QAnon \\
		& \emph{Sentiment} & Sentiment, where sentiment scores are weighted over all posts and comments by length. \\
		& \emph{Toxicity} & Average toxicity by user \\
		& \emph{Severe toxicity} & Average severe toxicity by user \\
		& \emph{Identity attack} & Average level of identity attacks by user \\
		& \emph{Insult} & Average level of insults by user \\
		& \emph{Profanity} & Average level of profanities by user \\
		& \emph{Threat} & Average level of threat by user \\
		& \emph{LIWC features} & Average LIWC scores by user \\
		& \emph{Empath features} & Average Empath scores by user \\
		\midrule
		\multirowcell{4}[0pt][l]{Network \\ Features} & \emph{Betweenness} & Betweenness centrality of user \\
		& \emph{Eigen} & Eigenvector centrality of user \\
		& \emph{In-degree} & In-degree centrality of user\\
		& \emph{Out-degree} & Out-degree centrality of user\\
		\midrule
		Content Features & \emph{Embeddings} & Text embeddings by user\\
		\bottomrule
		\multicolumn{3}{p{0.95\linewidth}}{$^1$ Part-of-speech tags comprise nouns, pronouns, adjectives, verbs, adpositions, and determiners, thus counting words in ``natural'' language.} 
	\end{tabular}
	\caption{List of features extracted per user.}
	\label{tbl:features}
\end{table*}

\subsection{Machine Learning Approach}
\label{sec:ml}

We use machine learning in order to discriminate QAnon from non-QAnon supporters based on the above features. For this, we use extreme gradient boosting (XGBoost) \cite{Chen.2016}. The outcome variable $y$ is set to $y=1$ if a user is classified as a QAnon supporter, and $y=0$ otherwise. The choice of XGBoost is consistent with research performing user profiling in other settings (\eg, to predict verified status on Twitter \cite{Paul.2019}). In our analysis, we fit a classifier using the combination of all feature groups to evaluate the overall performance. To study the predictive power of the different feature groups individually, we further fit separate classifiers to the respective set of user features, linguistic features, network features, and content features. Finally, we evaluate the performance based on the area under the receiver operating characteristic curve (ROC~AUC).

Our original sample is heavily skewed towards non-QAnon supporters (see Tbl.~\ref{tbl:size_qanon}). Therefore, we randomly selected a balanced subsample of $n=62,084$ users. Before training, we split our data into a training set (80\,\%) and a hold-out set (20\,\%) for the model evaluation. XGBoost is tuned using 10-fold cross-validation in combination with a grid search (details are in our GitHub). 

\section{Results}

\subsection{Size of QAnon Community on Parler (RQ1)}

To answer \textbf{RQ1}, we classify users on Parler into self-reported QAnon supporters vs. others by matching user bios against an extensive list of keywords that are characteristic of QAnon. This allows us to quantify the size of the QAnon community on Parler. Reassuringly, we remind that we refer to all other users as ``non-QAnon supporters.'' The results are shown in Tbl.~\ref{tbl:size_qanon}. Out of all English-speaking users in our sample, we find that a large number of Parler users (34,913) are self-reported QAnon supporters. This amounts to 5.5\,\% of all users, thus providing evidence that Parler hosts a comparatively large QAnon community. 

In addition, we provide summary statistics on the number of posts and comments associated with QAnon supporters on Parler (see Tbl.~\ref{tbl:size_qanon}). We find that QAnon supporters on Parler shared approximately 21.55\,M posts and 3.65\,M comments. This accounts for 14\,\%, and 9\,\%, respectively, of the overall posts and comments shared by users in our sample. Hence, QAnon supporters share proportionately more posts and comments than non-QAnon supporters on Parler. In the following, we will further study how user characteristics of QAnon and non-QAnon supporters on Parler differ.

\begin{table}[h!]
	\small
	\centering
	\linespread{0.5}
	\setlength{\tabcolsep}{4pt}
	\begin{tabular}{p{0.19\linewidth} L{0.19\linewidth}L{0.2\linewidth}L{0.28\linewidth}}
		\toprule
		\textbf{Class Label} & \textbf{\#Users} \textbf{[\%]} & \textbf{\#Posts} \textbf{[\%]} & \textbf{\#Comments} \textbf{[\%]} \\
		\midrule
		\multirow{2}{*}{QAnon} & 34,913 & 21,532,766 & 3,619,857  \\
        & [5.5\,\%] & [14\,\%] & [9\,\%] \\
		\multirow{2}{*}{Non-QAnon} &  603,952  & 136,316,477  & 38,701,991 \\
        & [94.5\,\%] & [86\,\%] & [91\,\%] \\
		\midrule 
		\multirow{2}{*}{Overall} & 638,865 & 157,849,243 & 42,321,848 \\
        & [100\,\%] & [100\,\%] & [100\,\%] \\
		\bottomrule
	\end{tabular}
	\caption{Descriptives summarizing the QAnon community on Parler.}
	\label{tbl:size_qanon}
\end{table}

\subsection{Comparison of User Characteristics between QAnon vs. non-QAnon Supporters (RQ2)}

To answer \textbf{RQ2}, we now compare QAnon vs. non-QAnon supporters on Parler across different user characteristics. For this purpose, we analyze descriptive statistics with respect to the extracted user features, linguistic features, network features, and content features.

\begin{figure*}[ht]
	\centering
	\captionsetup[subfigure]{position=top, justification=centering}
	\begin{subfigure}[b]{0.19\linewidth}
		\centering
		\caption{Posts}
		\includegraphics[width=\linewidth]{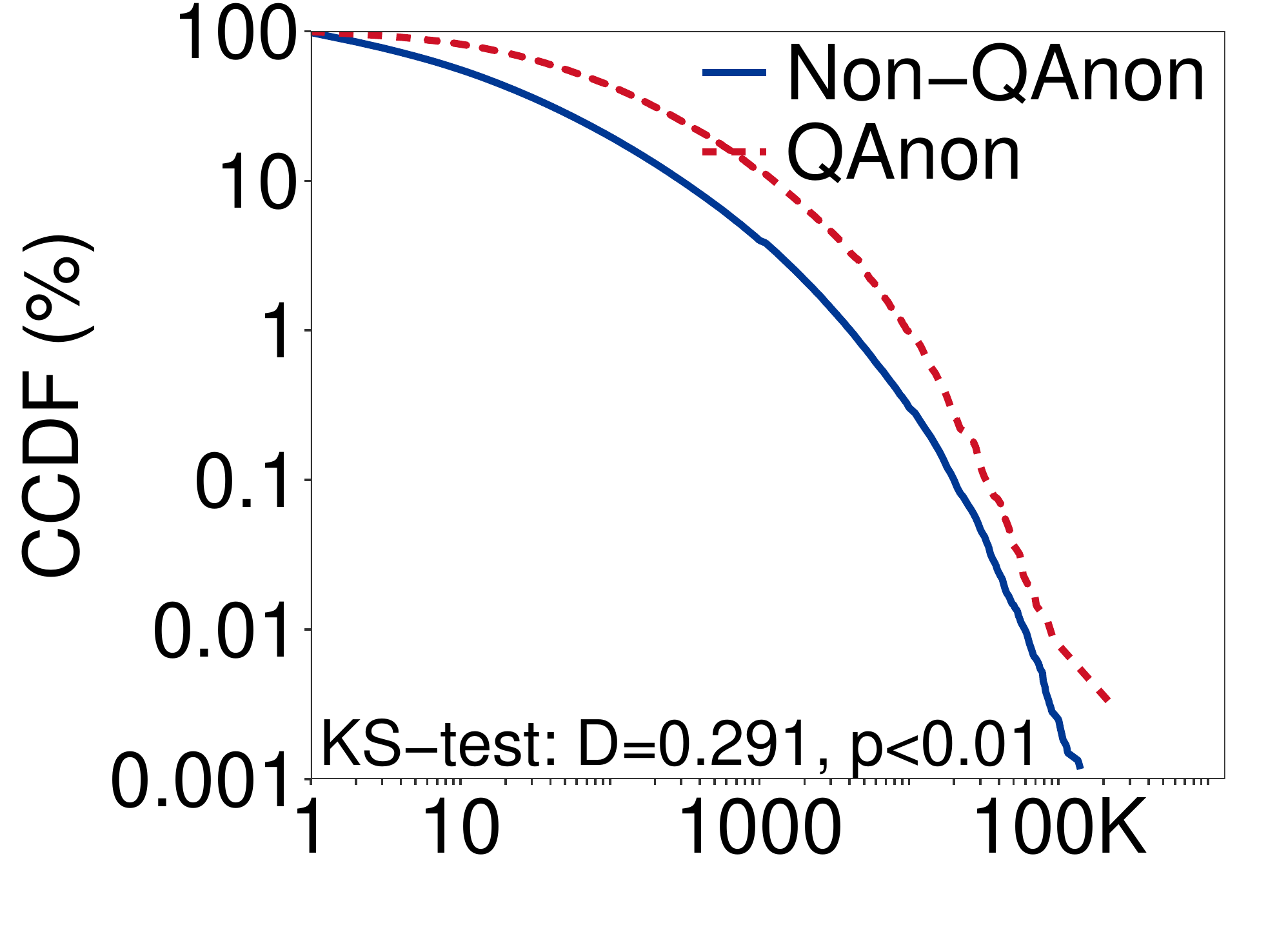}
		\label{fig:ccdf_posts}
	\end{subfigure}
	\begin{subfigure}[b]{0.19\linewidth}
		\centering
		\caption{Comments}
		\includegraphics[width=\linewidth]{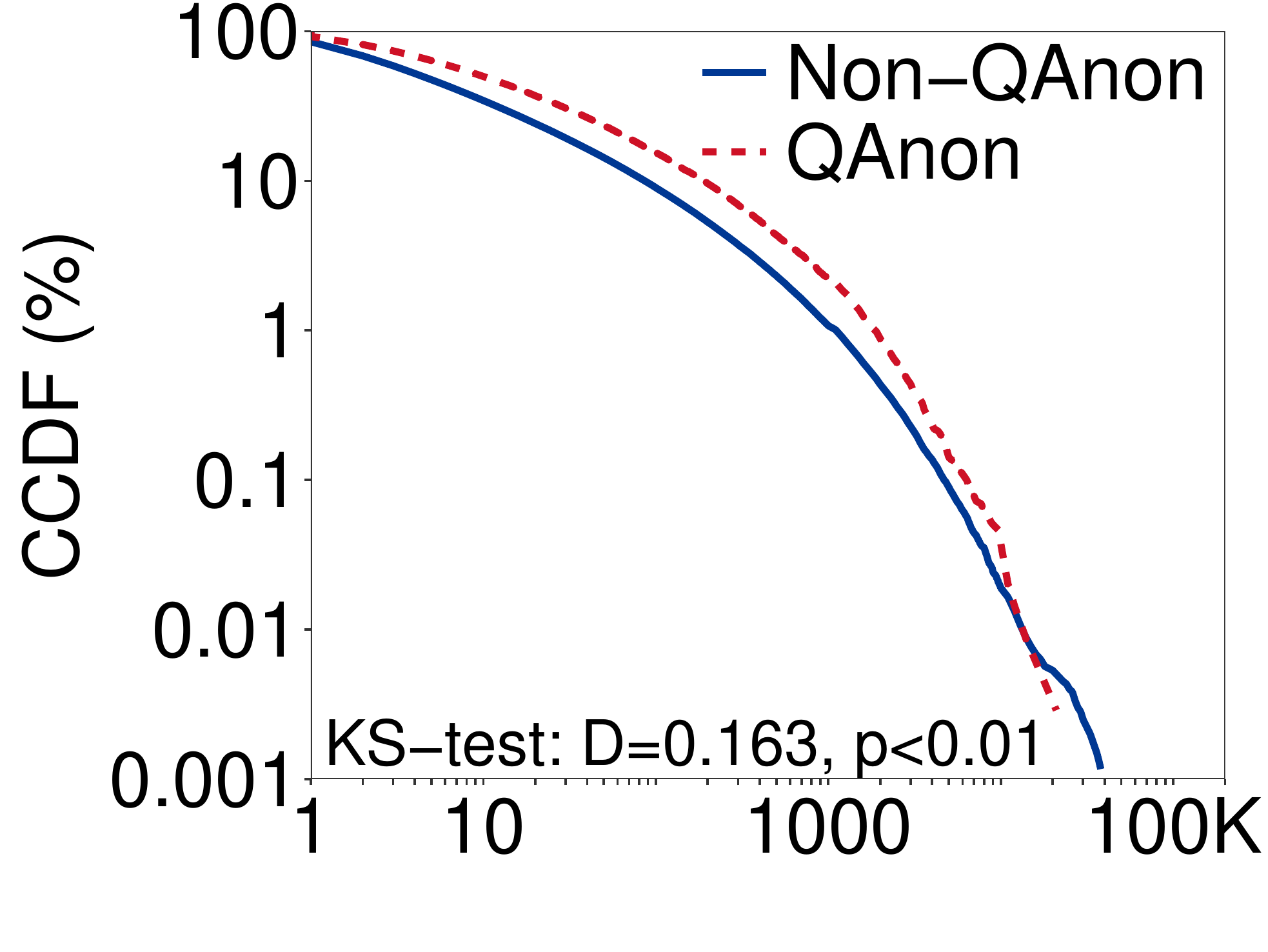}
		\label{fig:ccdf_comments}
	\end{subfigure}
	\begin{subfigure}[b]{0.19\linewidth}
		\centering
		\caption{Impressions (post)}
		\includegraphics[width=\linewidth]{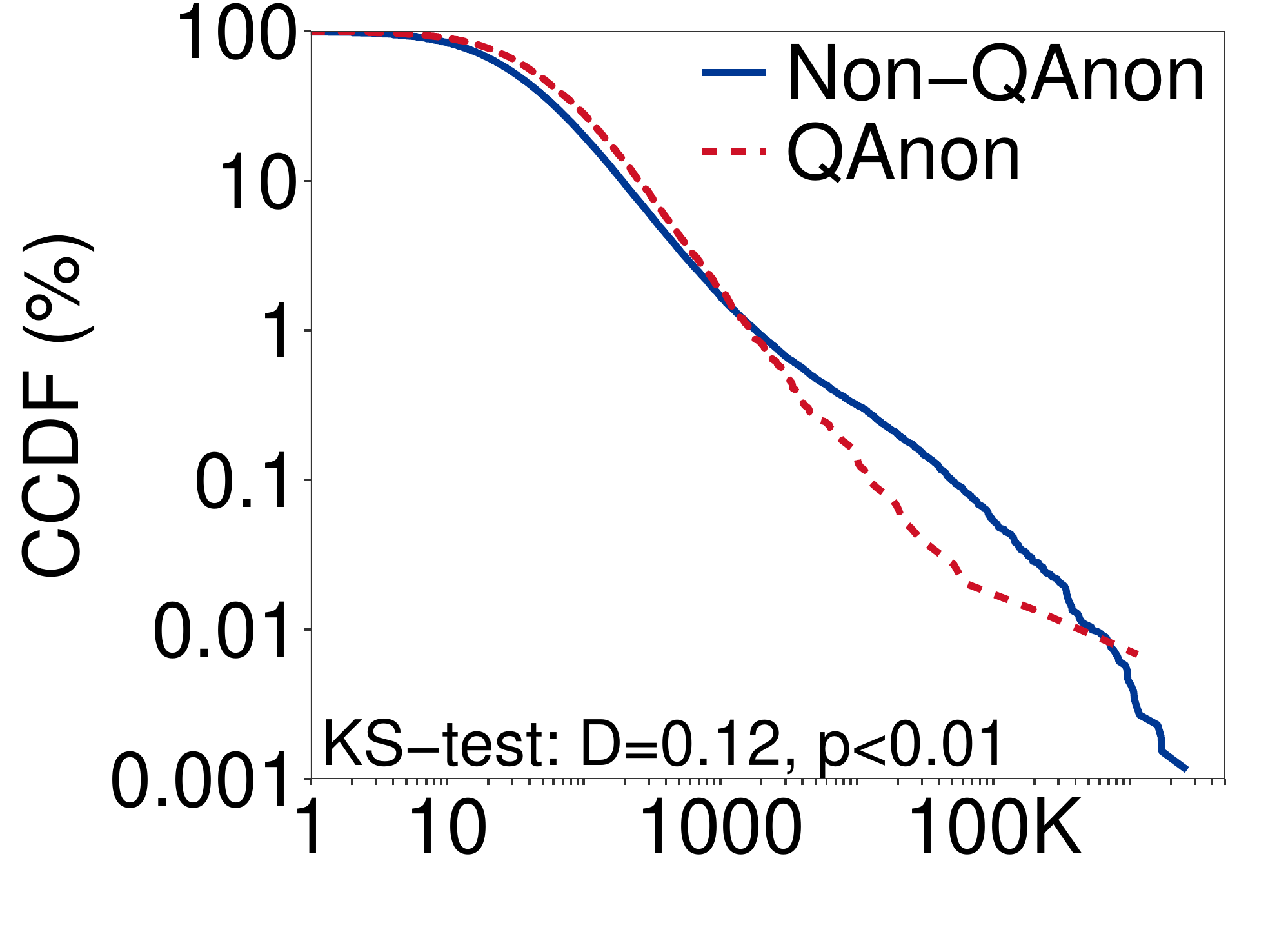}
		\label{fig:ccdf_impressions_posts}
	\end{subfigure}
	\begin{subfigure}[b]{0.19\linewidth}
		\centering
		\caption{Upvotes (post)}
		\includegraphics[width=\linewidth]{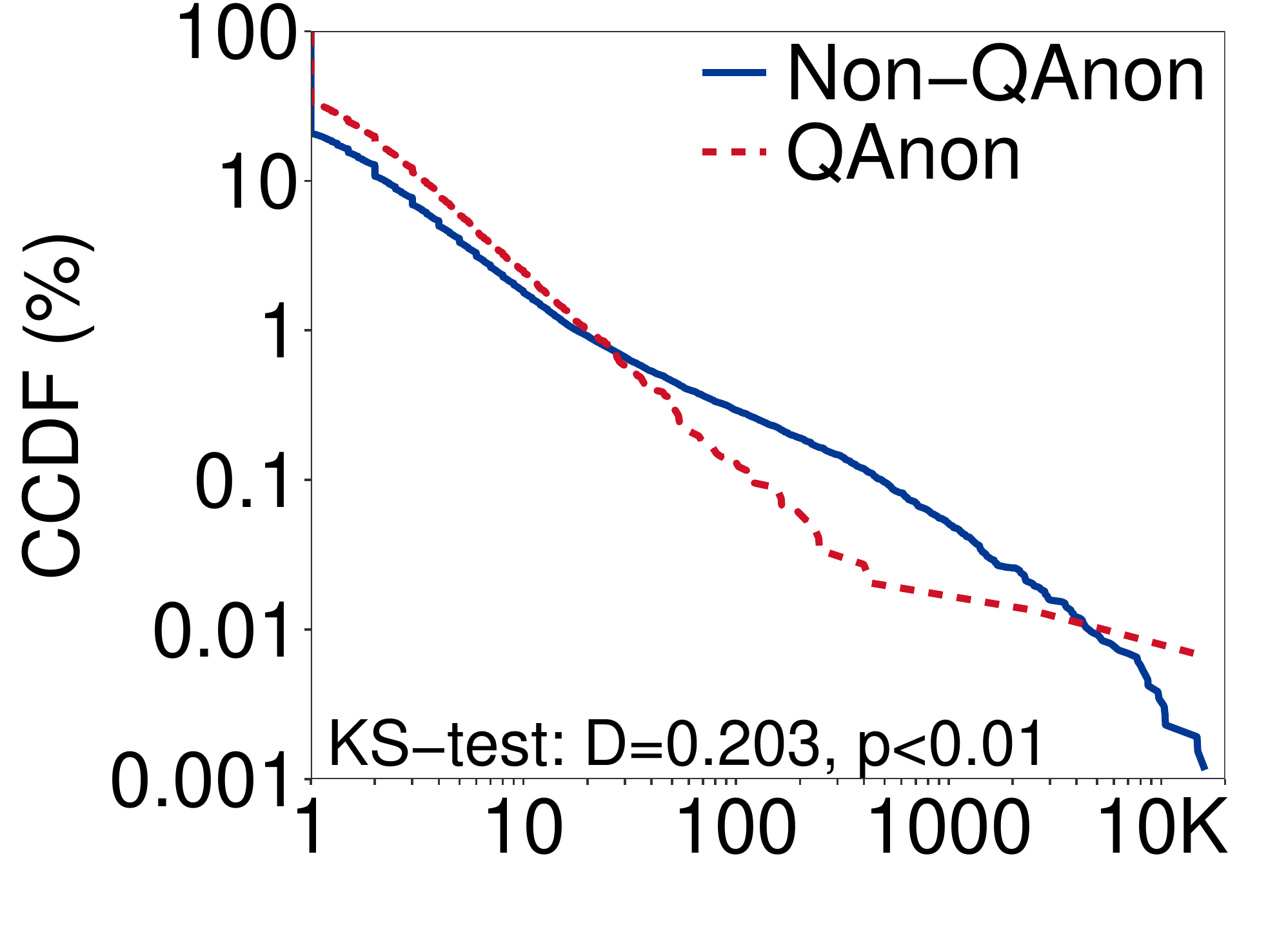}
		\label{fig:ccdf_upvotes_posts}
	\end{subfigure}
	\begin{subfigure}[b]{0.19\linewidth}
		\centering
		\caption{Upvotes (comment)}
		\includegraphics[width=\linewidth]{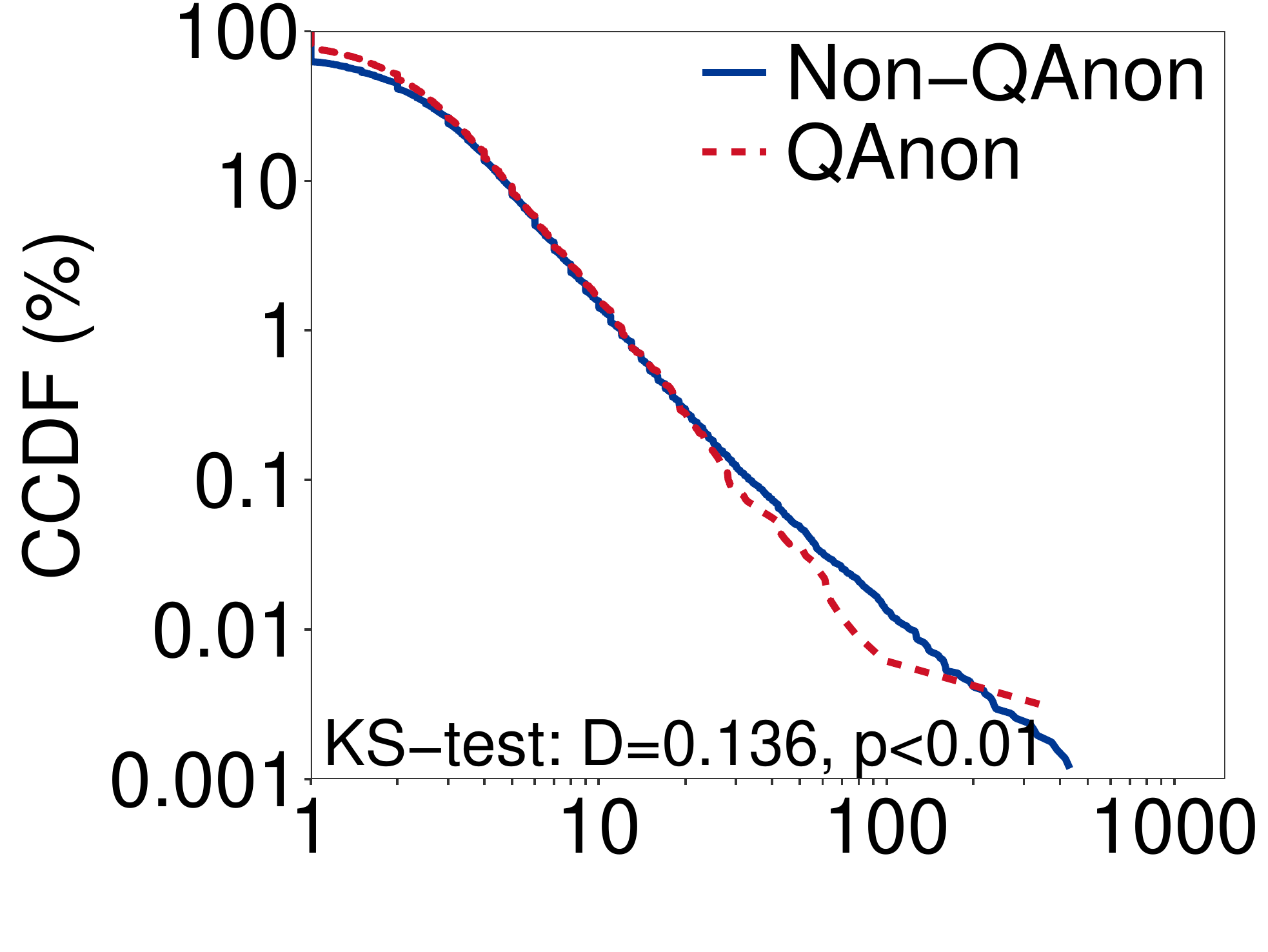}
		\label{fig:ccdf_upvotes_comments}
	\end{subfigure}
 
	\begin{subfigure}[b]{0.19\linewidth}
		\centering
		\caption{Downvotes (comment)}
		\includegraphics[width=\linewidth]{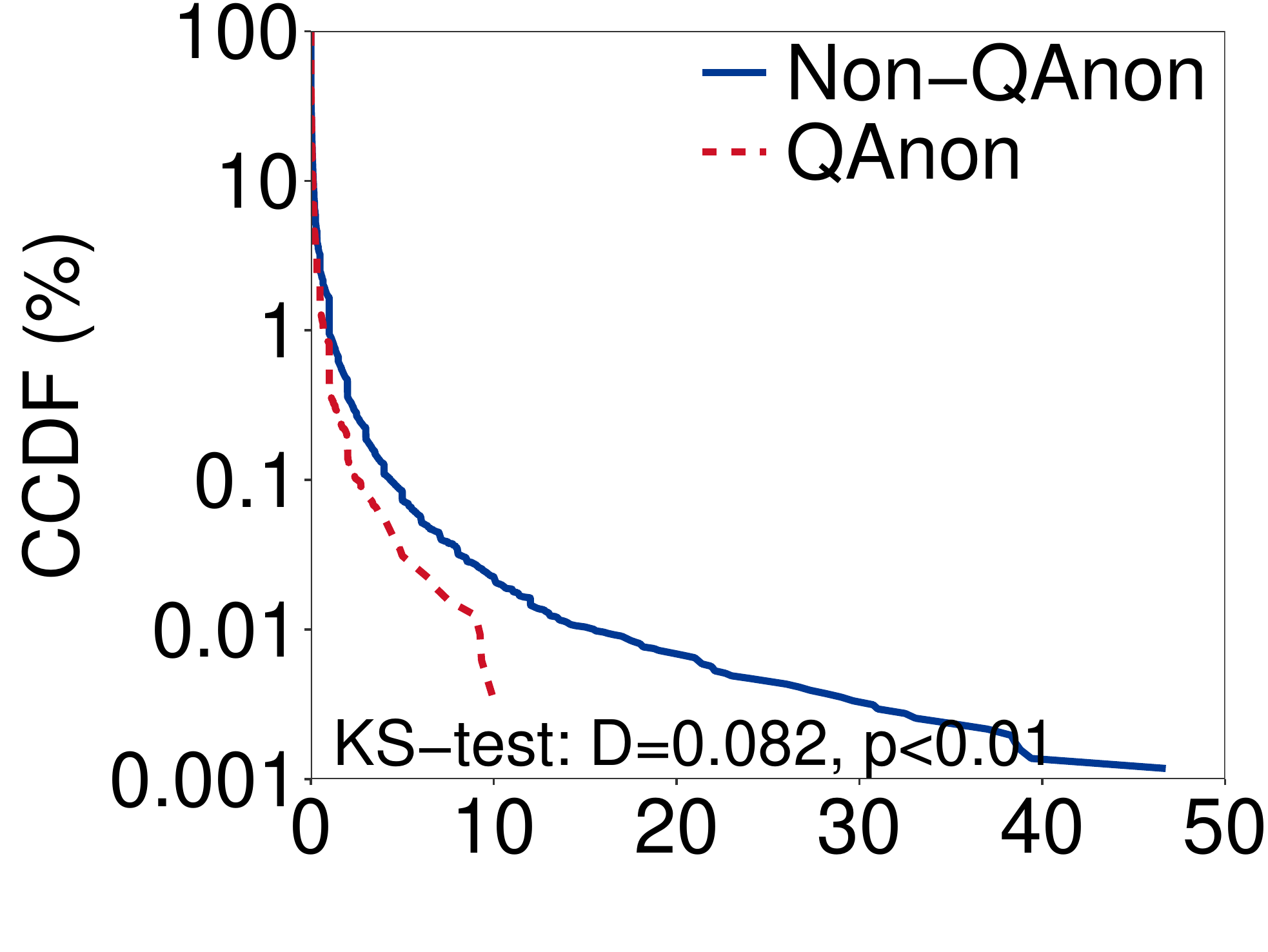}
		\label{fig:ccdf_downvotes_comments}
	\end{subfigure}
	\begin{subfigure}[b]{0.19\linewidth}
		\centering
		\caption{Handles}
		\includegraphics[width=\linewidth]{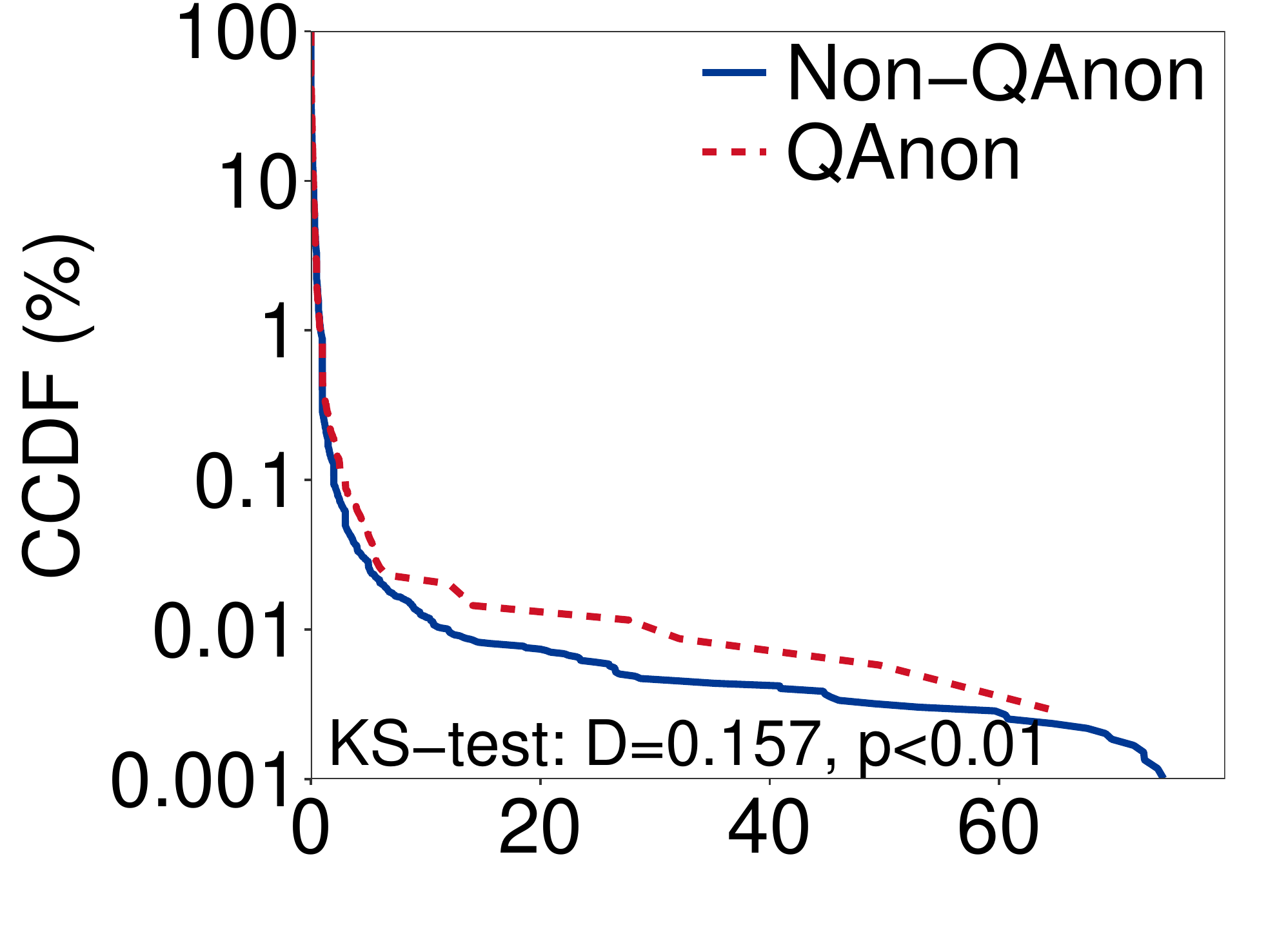}
		\label{fig:ccdf_handle}
	\end{subfigure}
	\begin{subfigure}[b]{0.19\linewidth}
		\centering
		\caption{Hashtags}
		\includegraphics[width=\linewidth]{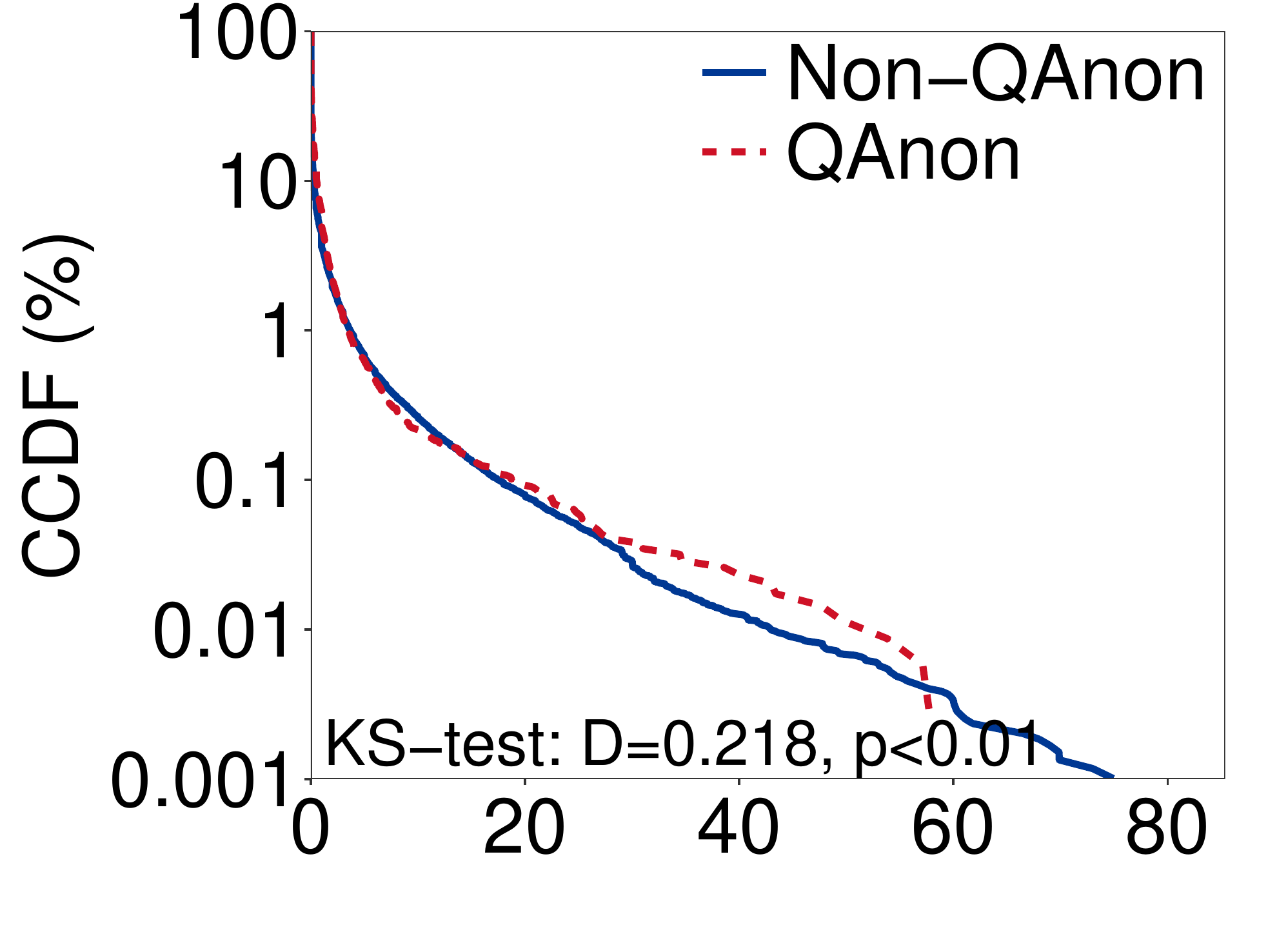}
		\label{fig:ccdf_hash}
	\end{subfigure}
	\begin{subfigure}[b]{0.19\linewidth}
		\centering
		\caption{URLs}
		\includegraphics[width=\linewidth]{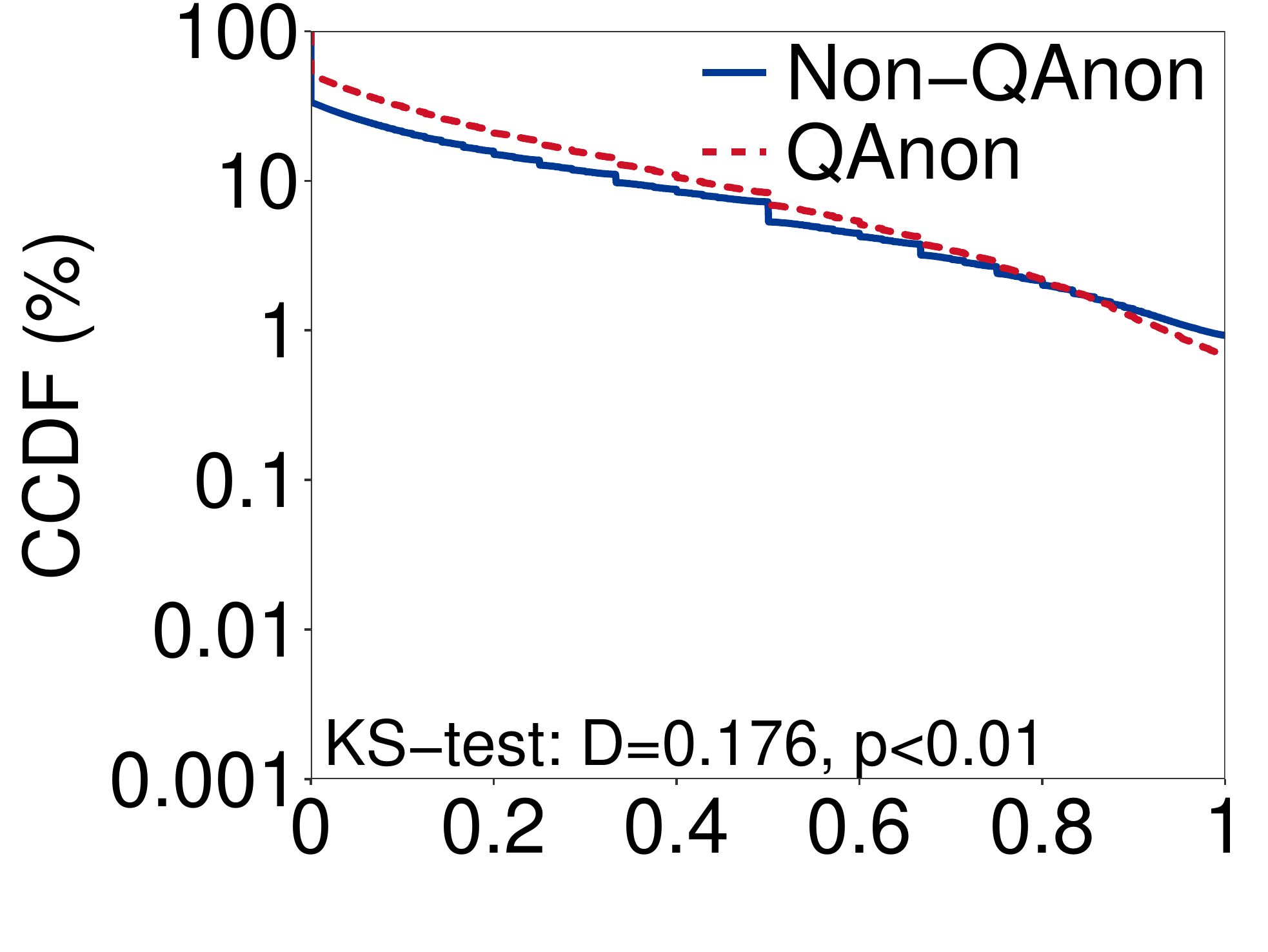}
		\label{fig:ccdf_urls}
	\end{subfigure}
	\begin{subfigure}[b]{0.19\linewidth}
		\centering
		\caption{Sentiment}
		\includegraphics[width=\linewidth]{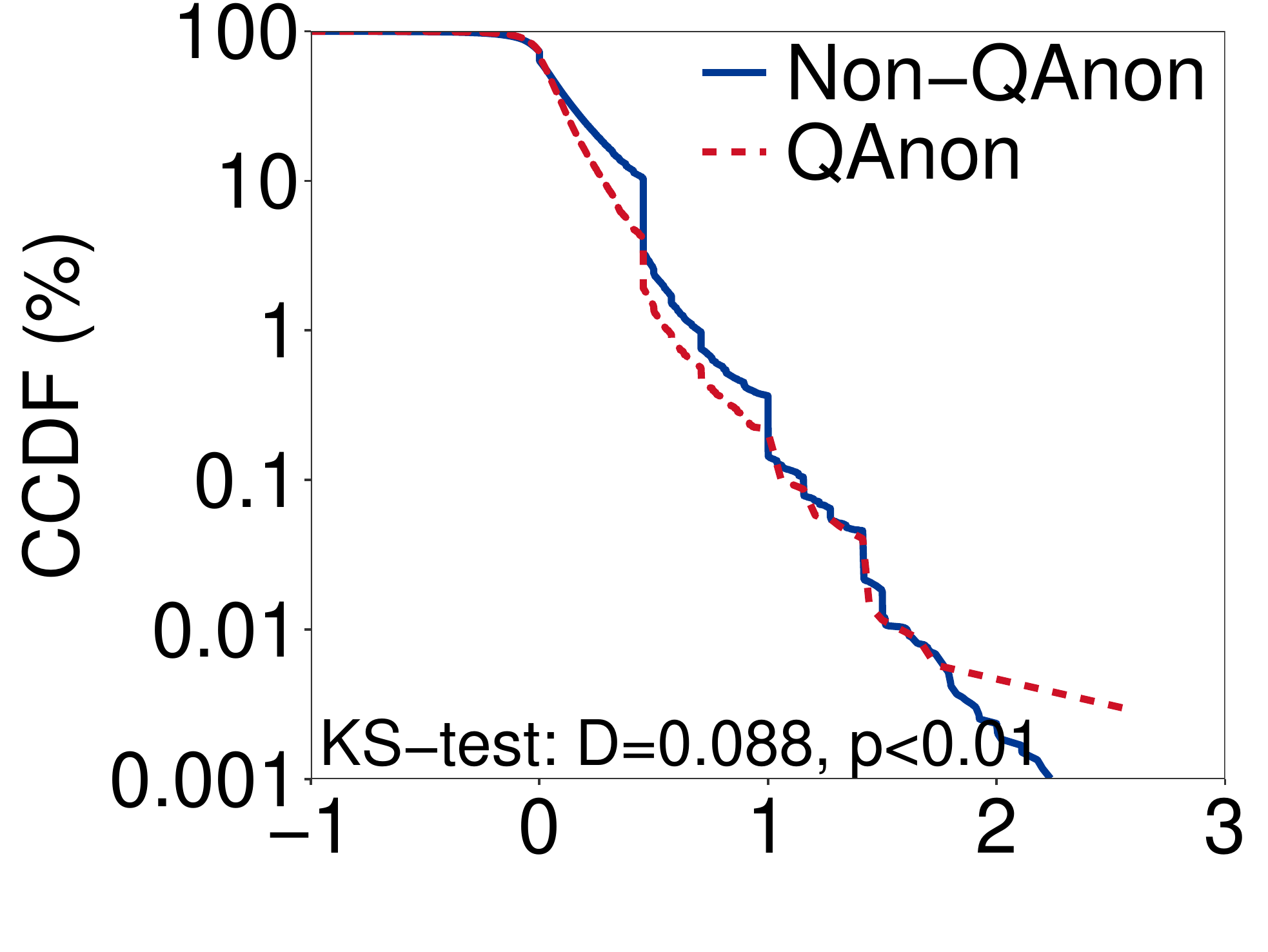}
		\label{fig:ccdf_sentiment}
	\end{subfigure}
 
	\begin{subfigure}[b]{0.19\linewidth}
		\centering
		\caption{Toxicity}
		\includegraphics[width=\linewidth]{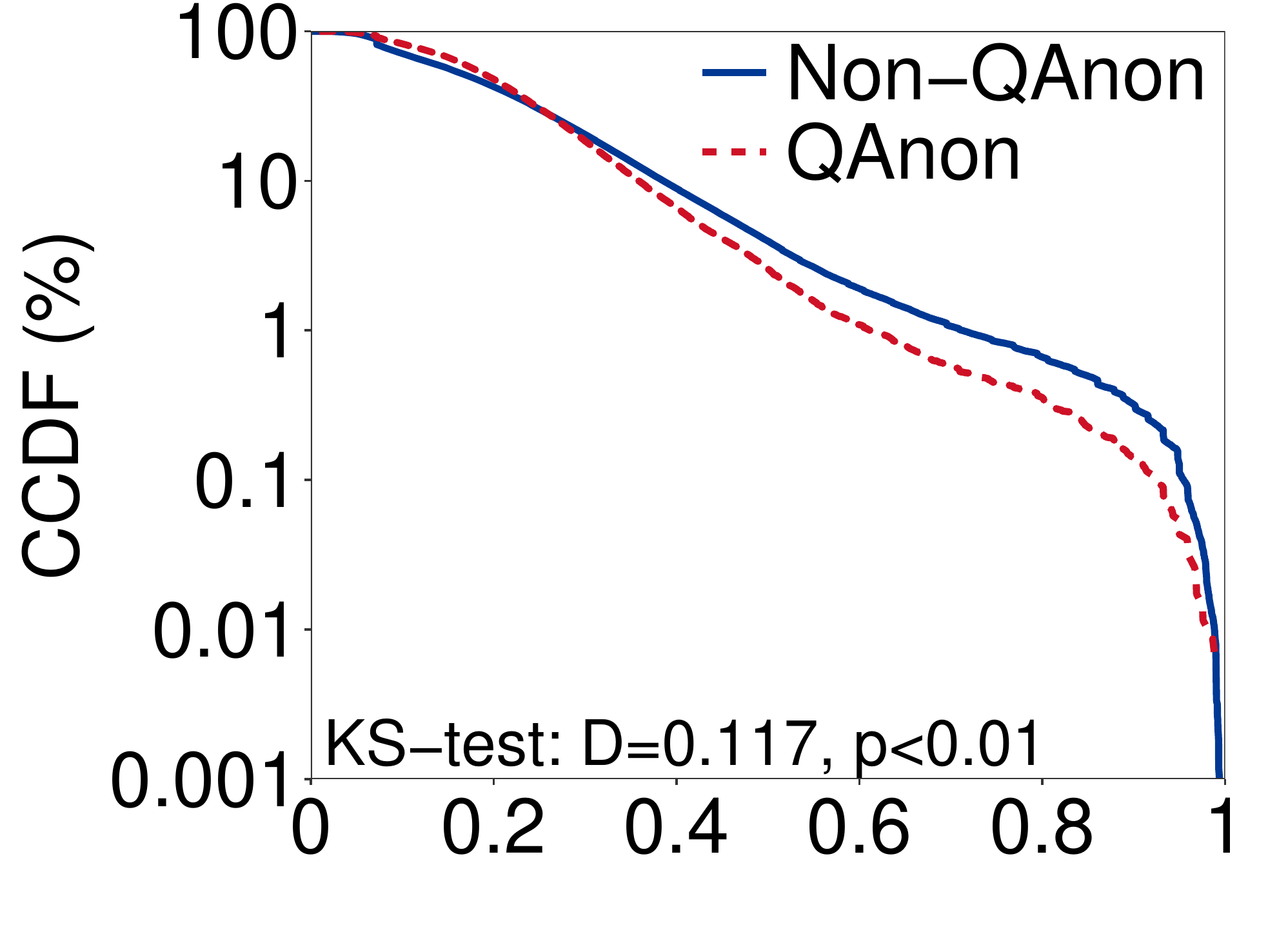}
		\label{fig:ccdf_toxicity}
	\end{subfigure}
	\begin{subfigure}[b]{0.19\linewidth}
		\centering
		\caption{Threat}
		\includegraphics[width=\linewidth]{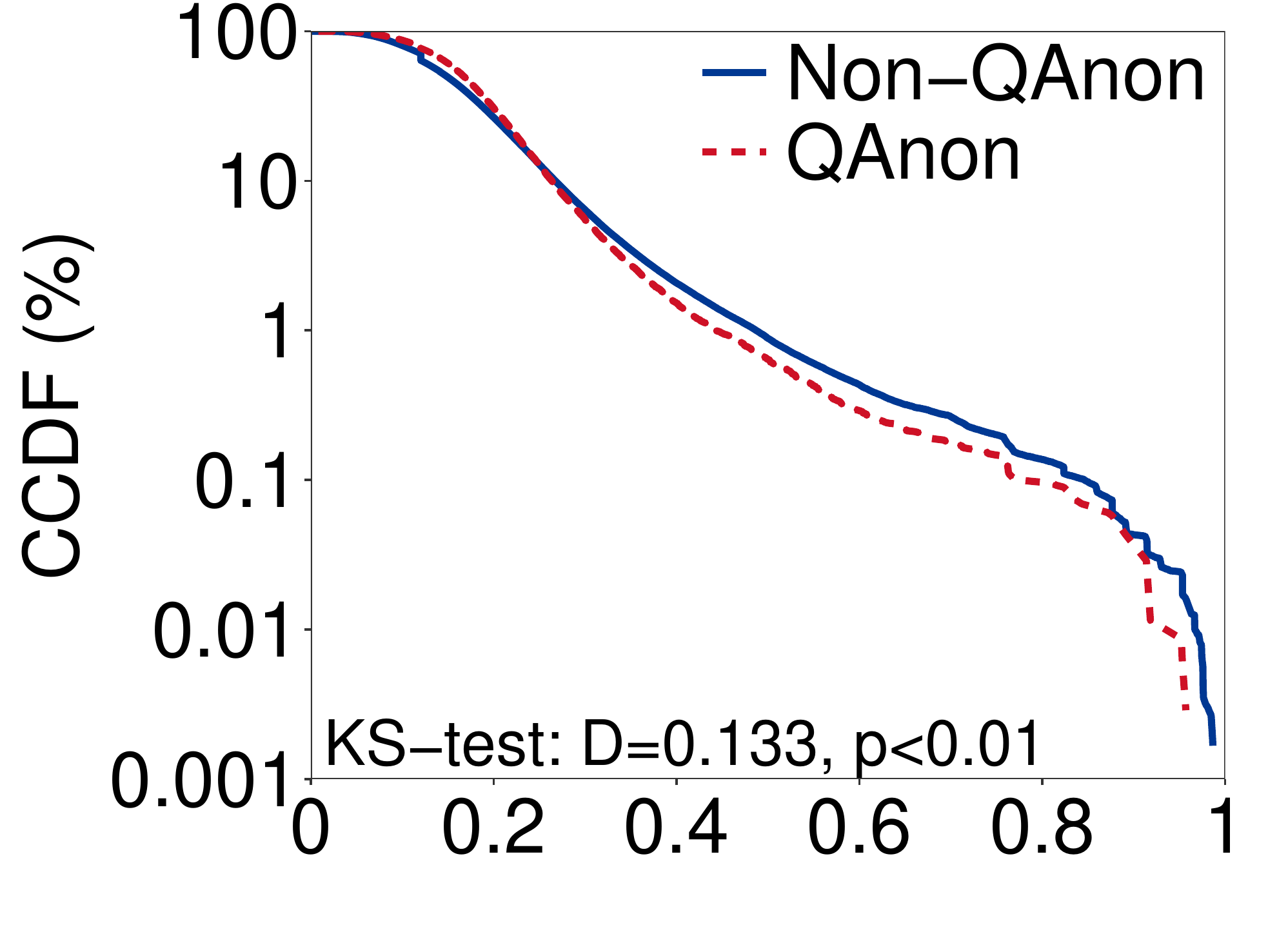}
		\label{fig:ccdf_threat}
	\end{subfigure}
	\begin{subfigure}[b]{0.19\linewidth}
		\centering
		\caption{Profanity}
		\includegraphics[width=\linewidth]{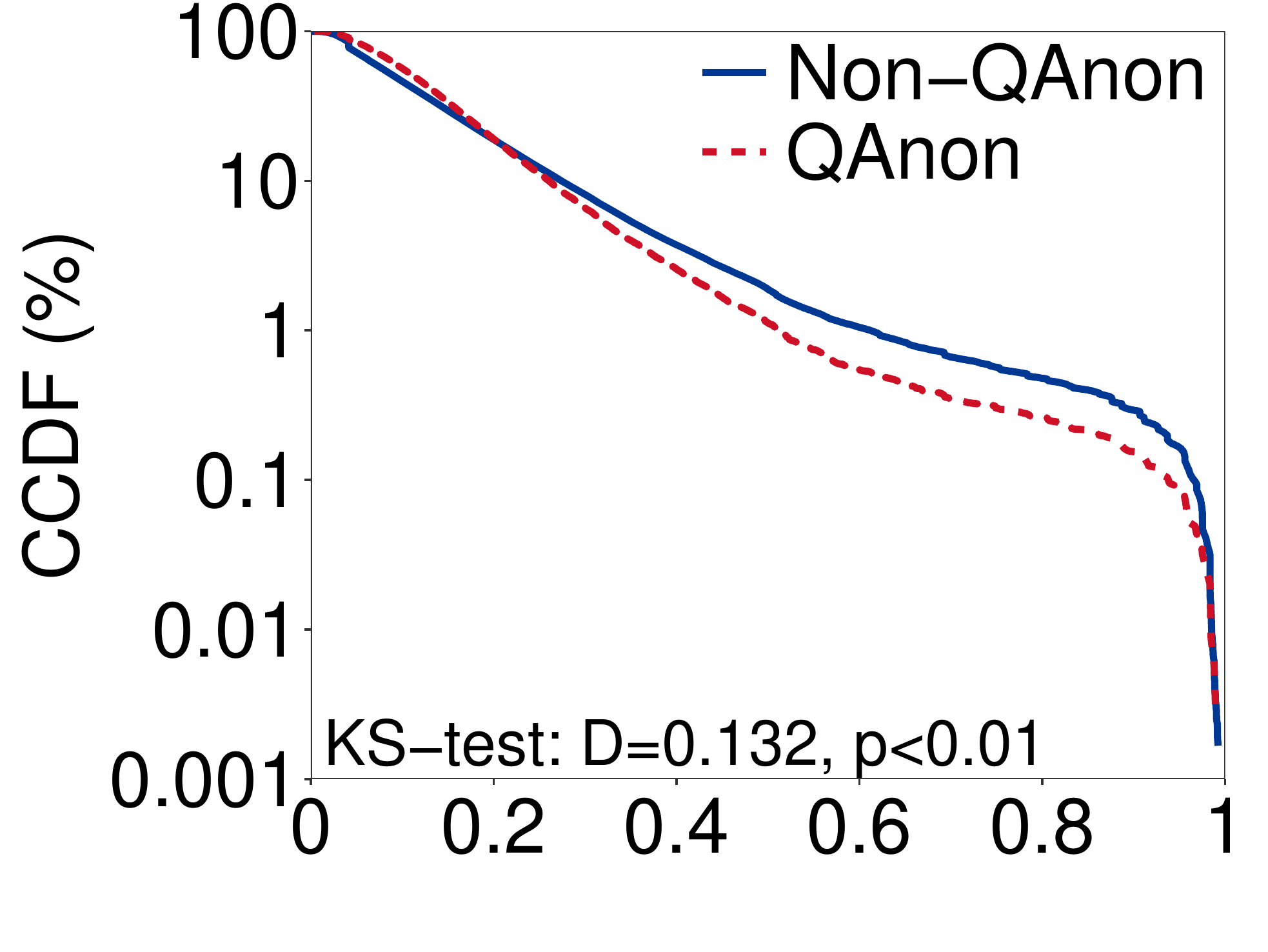}
		\label{fig:ccdf_profanity}
	\end{subfigure}
	\begin{subfigure}[b]{0.19\linewidth}
		\centering
		\caption{Followers}
		\includegraphics[width=\linewidth]{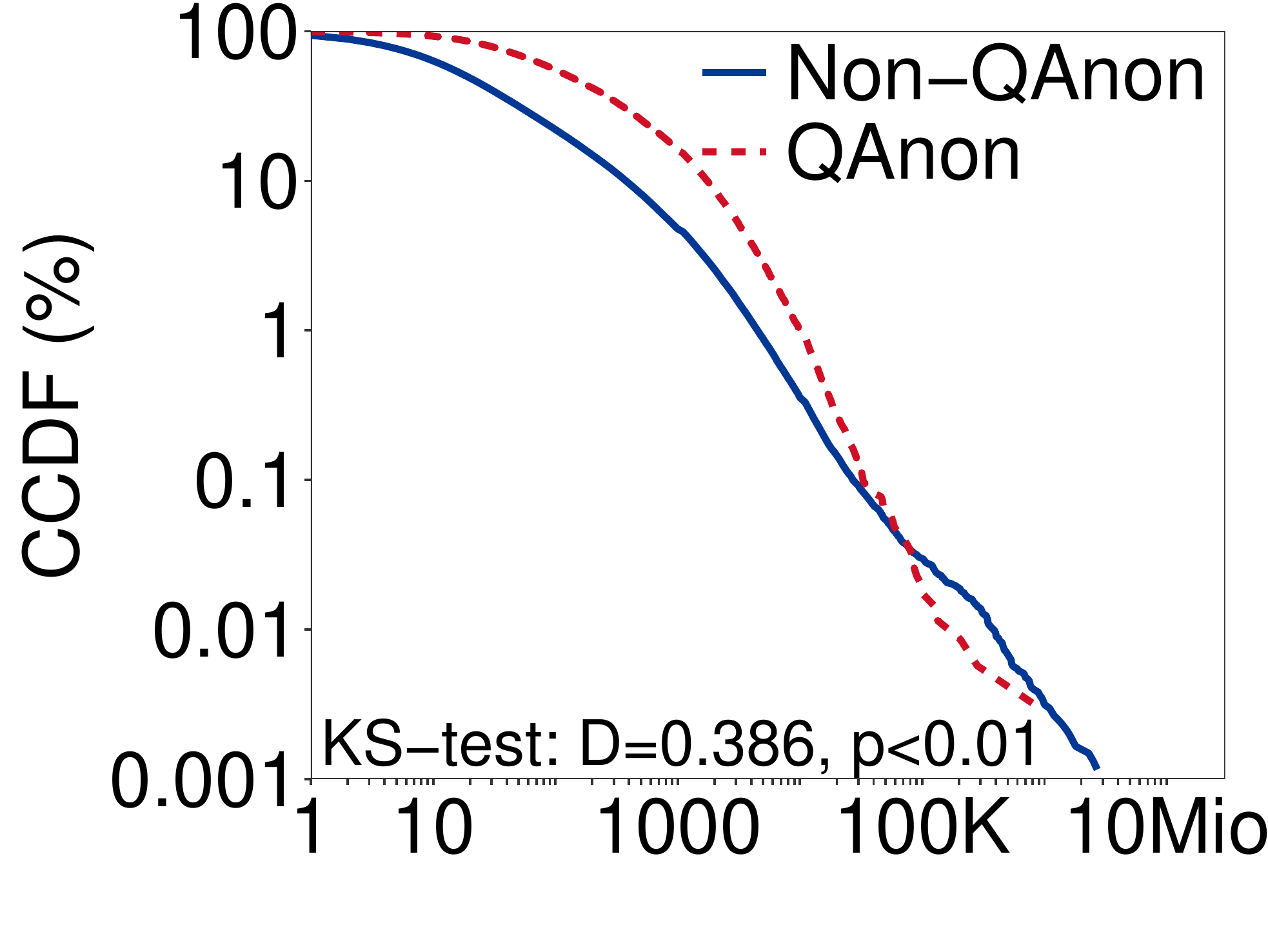}
		\label{fig:ccdf_followers}
	\end{subfigure}
	\begin{subfigure}[b]{0.19\linewidth}
		\centering
		\caption{Following}
		\includegraphics[width=\linewidth]{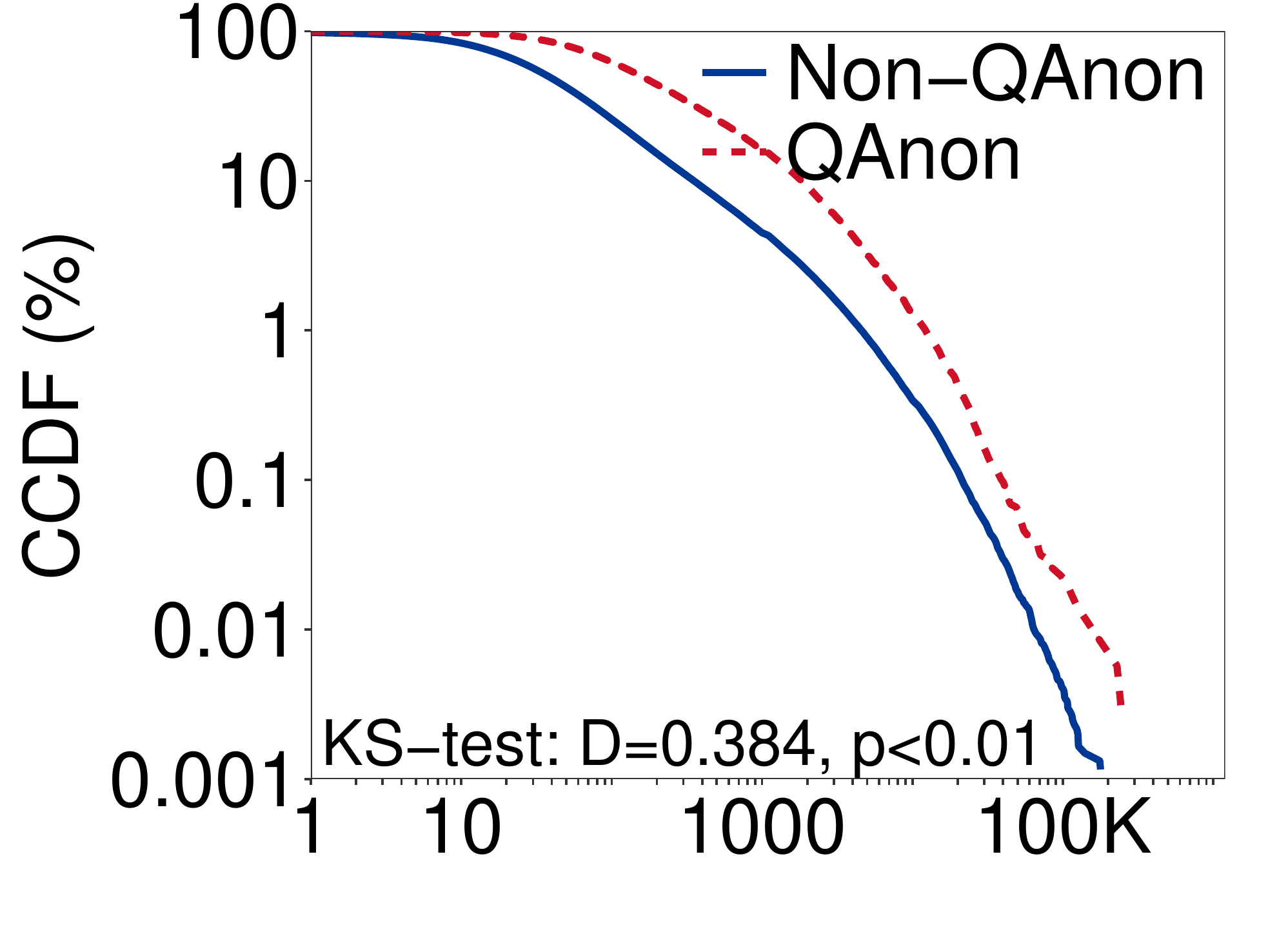}
		\label{fig:ccdf_followees}
	\end{subfigure}
	\caption{Complementary cumulative distribution functions (CCDFs) for selected features.}
	\label{fig:ccdfs}
\end{figure*}

\vspace{0.2cm}
\noindent
\emph{How active are QAnon supporters on Parler?}

QAnon supporters are a bigger threat if they are especially active. Abusive users tend to be more active on social media \cite{Ribeiro.2018}, which might also apply to QAnon users. Hence, we compare the number of posts and comments shared by QAnon vs. non-QAnon supporters on Parler. Specifically, we compute the complementary cumulative distribution function (CCDF) for both variables and test for statistically significant differences in the distributions using a Kolmgorov-Smirnov (KS) test \cite{Smirnov.1939}. We find that QAnon supporters share significantly more posts and comments compared to non-QAnon supporters ($p<0.01$) (see Fig.~\ref{fig:ccdfs}a,b). In particular, the average number of posts (comments) is 621.67 (104.51) for QAnon supporters vs. 227.04 (64.46) for non-supporters on Parler. Hence, QAnon users have a large impact on the content discussed on Parler.

\vspace{0.2cm}
\noindent
\emph{When did QAnon supporters join Parler?}

The timing of when QAnon supporters migrated to Parler is important, as it informs which user base -- Qanon supporters or non-supporters -- drives the growth of the platform. Hence, we compare the account age of QAnon and non-QAnon supporters. Overall, the majority of users joined only in a later stage of the platform ($\approx$ from June~2020 onward), which coincides, for example, with the 2020 U.S. presidential election and Twitter's increased efforts to manage content \cite{Aliapoulios.2021}. However, on average, QAnon supporters joined Parler earlier than non-QAnon supporters. The average account age for QAnon supporters is 266.25 days (interquartile range [IQR]: 193.85 to 248.84), whereas, for non-QAnon supporters, it is 203.60 days (interquartile range [IQR]: 84.14 to 221.68) and thus around 2 months earlier. Overall, this suggests that QAnon supporters were early adopters of so-called ``free speech'' platforms such as Parler (\eg, to openly discuss the conspiracy).

\vspace{0.2cm}
\noindent
\emph{How popular is content from QAnon supporters on Parler?}

We now examine the virality of QAnon-related content. In the past, QAnon-related content has gone viral on mainstream social media \cite{Sternisko.2020}. As such, we expect content by QAnon supporters to be more popular on Parler compared to content shared by non-supporters. To check this, we compare the popularity of posts and comments shared by QAnon vs. non-QAnon supporters on Parler. In doing so, we compare the number of upvotes per post, upvotes per comment, downvotes per comment, and impressions per user. We observe mixed patterns regarding the popularity of the posted content on Parler. Posts from both groups are, on average, almost equally likely to be upvoted (QAnon: 3.00 vs. non-QAnon: 3.62) (see Fig.~\ref{fig:ccdf_upvotes_posts}). Similarly, comments by QAnon supporters receive on average a comparable number of upvotes (QAnon: 2.53 vs. non-QAnon: 2.27) (see Fig.~\ref{fig:ccdf_upvotes_comments}). Nevertheless, a KS-test for each of the variables suggests statistically significant distributional differences for all variables ($p<0.01$). In contrast, posts by non-QAnon supporters on average receive substantially less impressions (see Fig.~\ref{fig:ccdf_impressions_posts}). In particular, non-QAnon supporters on average receive 406.21 impressions, while, for QAnon supporters, the average number of impressions is 249.09. Even though there is more QAnon-related content on Parler, the posts are less viral and far-reaching (as opposed to non-QAnon-related content).

\vspace{0.2cm}
\noindent
\emph{How do QAnon share content on Parler?}

QAnon supporters have developed specific hashtags \cite{Sharma.2022} and collectively investigate the cabal and decipher Q drops \cite{Aliapoulios.2022}. This might relate to an extensive use of hashtags, handles, and URLs. Thus, we now compare \emph{``how''} content by QAnon and non-QAnon supporters differs on Parler. We find that QAnon supporters use more handles than non-QAnon supporters (see Fig.~\ref{fig:ccdf_handle}). Evidently, QAnon supporters also use more hashtags and share URLs more frequently than non-QAnon supporters (see Fig.~\ref{fig:ccdf_hash} and Fig.~\ref{fig:ccdf_urls}, respectively). The distributional difference is statistically significant for all variables ($p<0.01$). Overall, this suggests that QAnon supporters make strategic use of handles, hashtags, and URLs. A possible explanation is that this might help to actively discuss the conspiracy with other supporters, reach certain audiences, and share external content to sources of conspiratorial materials.

\vspace{0.2cm}
\noindent
\emph{How does sentiment, toxicity, threat, and profanity vary between QAnon and non-QAnon supporters?}

On the one hand, the QAnon narrative (and many theories around cabals, sex trafficking, etc.) lets one expect high levels of toxicity, threat, and profanity. On the other hand, observations from other alt-right social media platforms such as Gab, 4chan, and 8kun suggest lower levels of toxicity, threat, and profanity for posts authored by QAnon supporters compared to other content on those platforms \cite{Papasavva.2021, Aliapoulios.2022}. Hence, we study differences in sentiment, toxicity, threat, and profanity. For all variables, we find distributional differences at statistically significant levels (see Fig.~\ref{fig:ccdf_sentiment}--\ref{fig:ccdf_profanity}). QAnon supporters appear to use more positive sentiment compared to non-QAnon supporters. Furthermore, we find that posts and comments by QAnon supporters express lower levels of toxicity, threat, and profanity. Overall, this resembles the general narrative of QAnon pointing to a ``brighter'' future after the cabal is defeated \cite{Zuckerman.2019}.

\vspace{0.2cm}
\noindent
\emph{How are (non-)QAnon supporters connected on Parler?}

We compare the network structure of QAnon vs. non-QAnon supporters on Parler by analyzing (i)~the friendship network and (ii)~the repost network:

\textbf{(i)~Friendship Network:} Abusive users on social media tend to maintain larger friendship networks \cite{Ribeiro.2018}. QAnon users might also build a large friendship network to reach other supporters or convince non-supporters. Hence, we compare the number of followers and followees of QAnon and non-QAnon supporters. QAnon supporters generally have more followers than non-QAnon supporters (see Fig.~\ref{fig:ccdf_followers}). The corresponding means amount to 799.20 (for QAnon supporters) vs. 416.25 (for non-QAnon supporters). In a similar vein, QAnon supporters are following more users than non-QAnon supporters (see Fig.~\ref{fig:ccdf_followees}). On average, QAnon supporters follow 856.54 users vs 272.44 users for non-QAnon supporters. For both variables, the distributional differences are statistically significant as suggested by a KS-test ($p<0.01$). This implies that QAnon supporters connect to a large number of users on Parler and thus attain an influential role in the network.

\begin{figure}
	\begin{subfigure}{0.49\linewidth}
		\centering
		\caption{Followers}
		\includegraphics[width=0.85\linewidth]{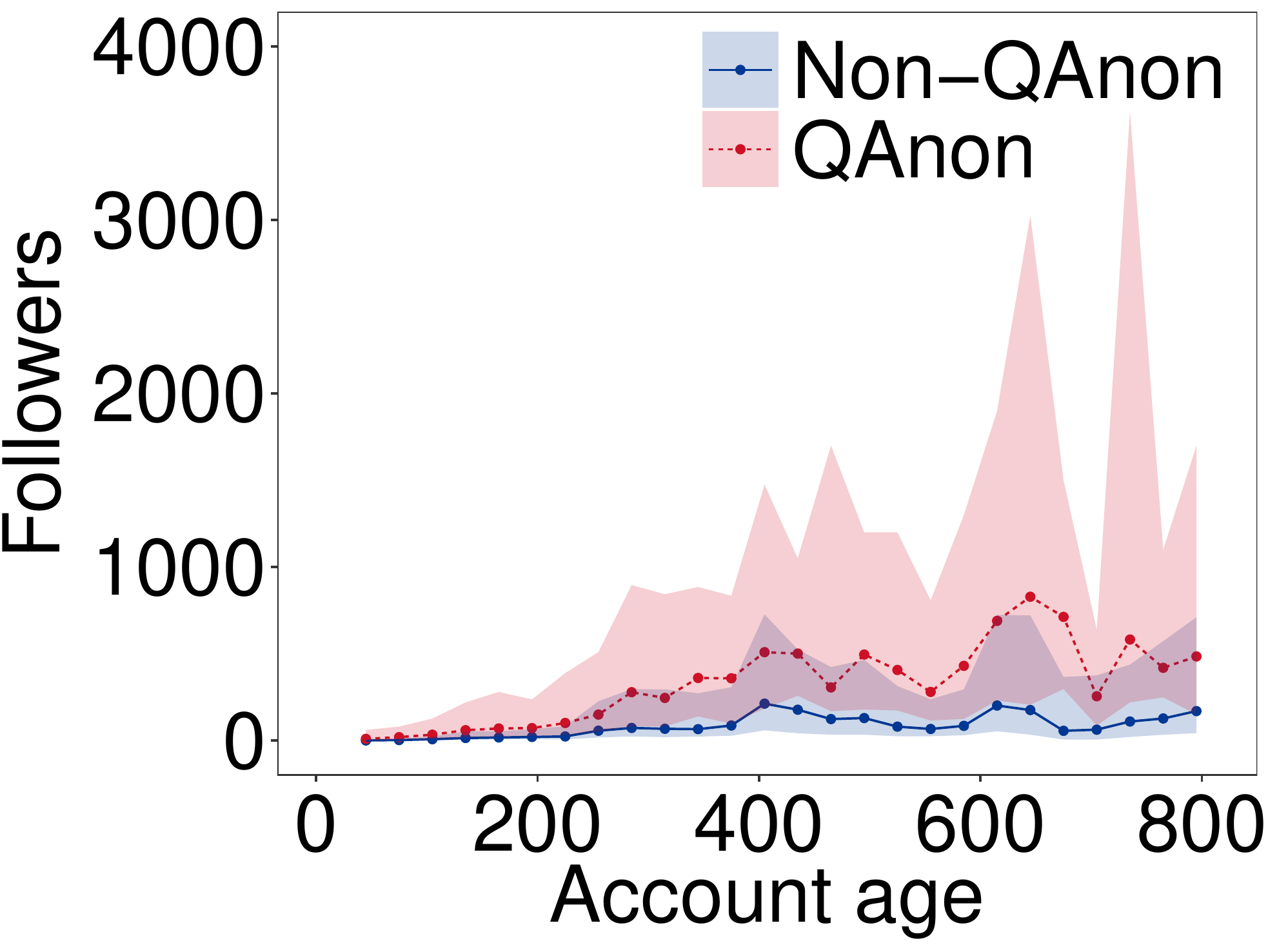}
		\label{fig:account_age_followers}
	\end{subfigure}
	\begin{subfigure}{0.49\linewidth}
	\centering
	\caption{Followees}
	\includegraphics[width=0.85\linewidth]{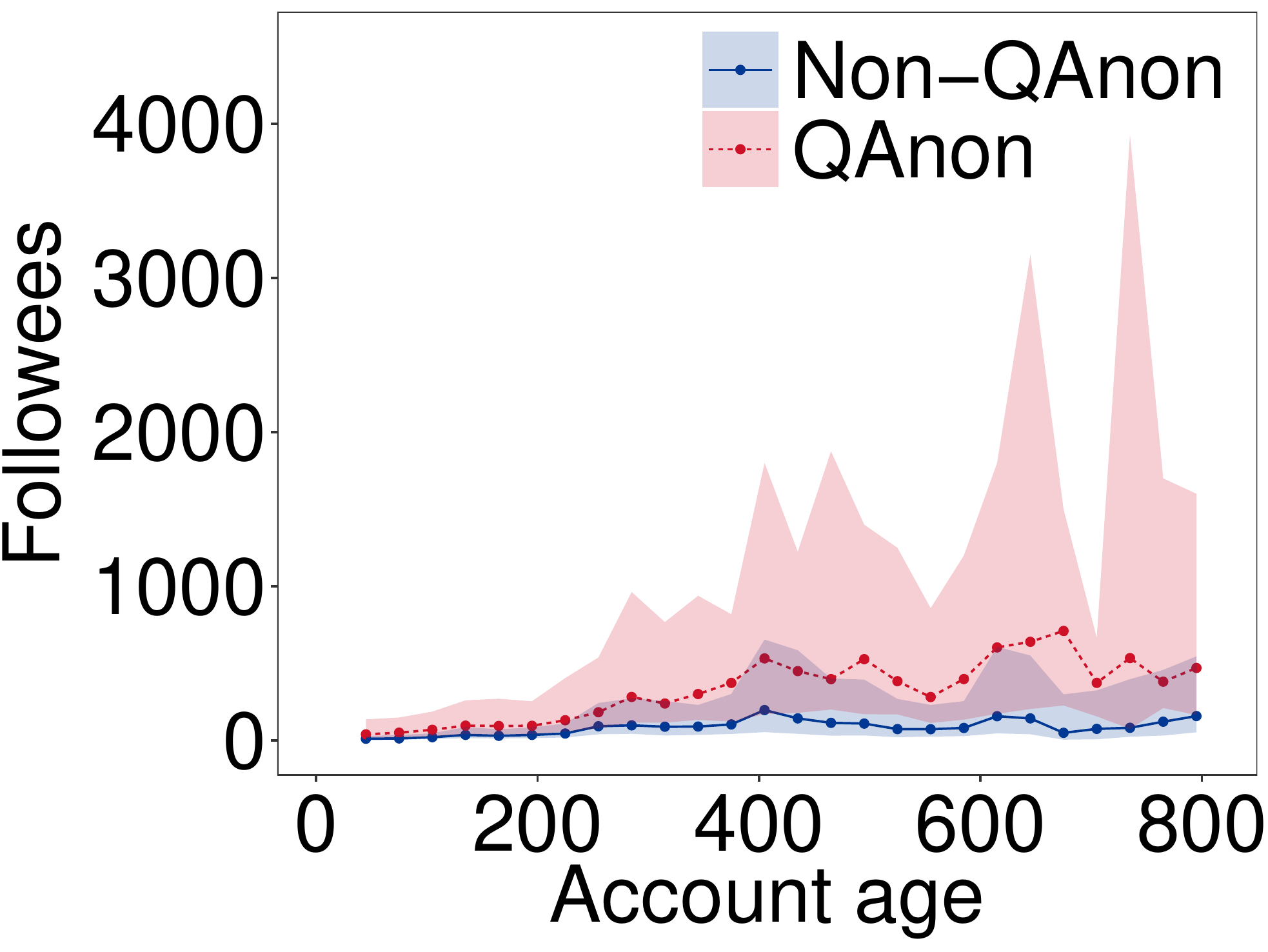}
	\label{fig:account_age_followees}
	\end{subfigure}
	\caption{Median number of (a)~followers and (b)~followees of QAnon and non-QAnon supporters by account age. Shading represents the interquartile range.}
	\label{fig:account_age_friendship}
\end{figure}

QAnon supporters, on average, joined Parler earlier compared to non-supporters. The additional time on the platform might have allowed QAnon supporters to grow a larger friendship network. Thus, we check how the size of the friendship network depends on the account age and compute the median number of followers and followees of QAnon supporters and non-QAnon supporters for 30-day intervals. The results are shown in Fig.~\ref{fig:account_age_friendship}. We find that the median number of followers is higher for QAnon supporters compared to non-QAnon supporters for each interval (see Fig.~\ref{fig:account_age_followers}). Similarly, QAnon supporters also follow more accounts compared to non-QAnon supporters regardless of their account age (see Fig.~\ref{fig:account_age_followees}). Overall, this shows how users grow on Paler and suggests that QAnon supporters quickly gain followers and followees on the platform.

\begin{figure}[ht]
	\centering
	\includegraphics[width=0.7\linewidth]{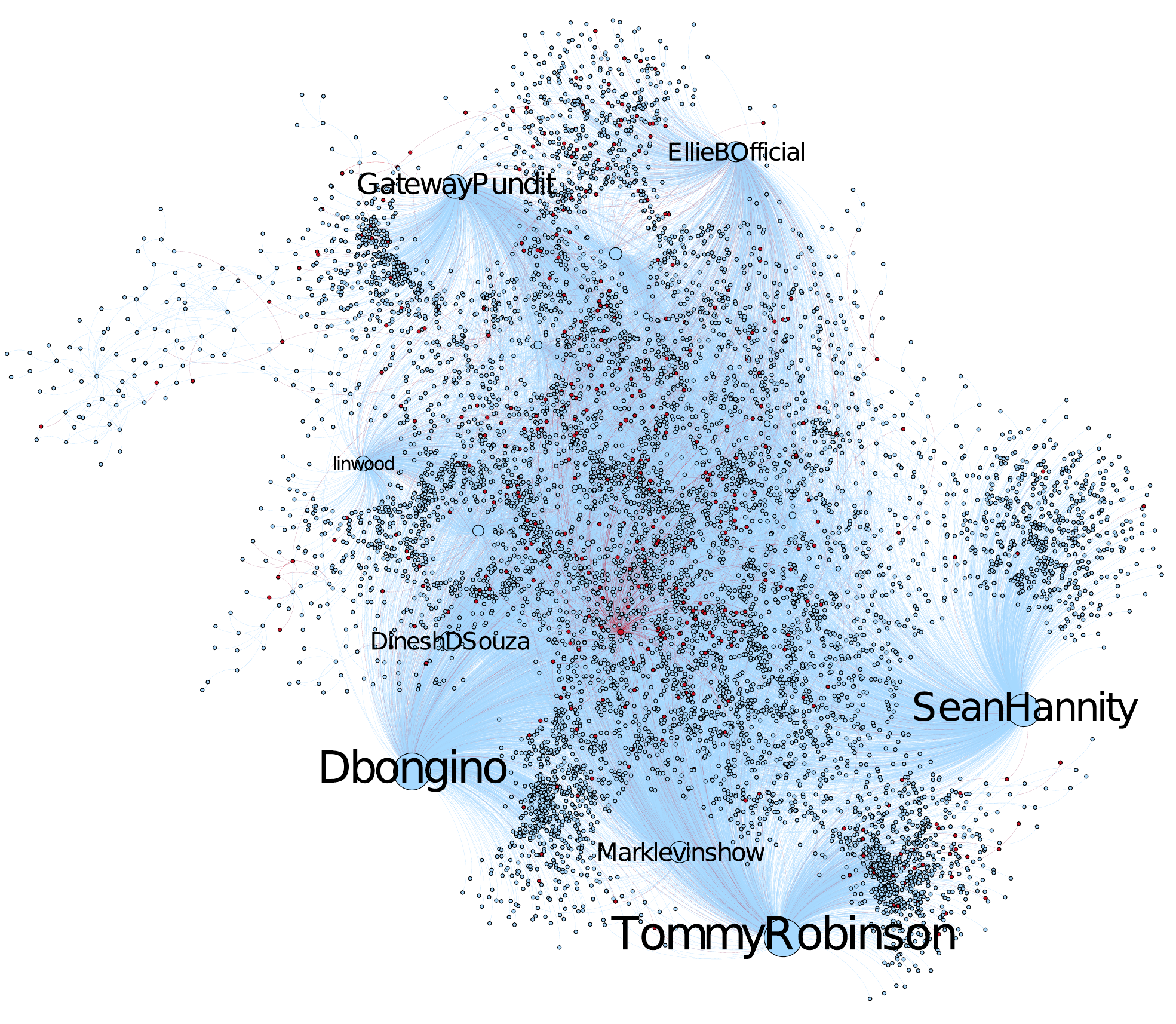}
	\caption{Social network plot of Parler showing user interactions between November 2020 and January 2021. QAnon supporters are colored in red. The size of the nodes varies by the overall number of interactions of a user (\ie, reposts and comments). For readability, only users who interacted at least 100 times are shown.}
	\label{fig:network_repost}
\end{figure}

\textbf{(ii)~Repost Network:} Content by QAnon supporters tends to receive substantially fewer impressions compared to non-supporters (see above). This implies that the repost network of QAnon and non-QAnon supporters on Parler differs. Fig.~\ref{fig:network_repost} illustrates the repost network for highly connected users on Parler between November 2020 and January 2021. We find that QAnon supporters (colored in red) take a central position in the Parler network. Furthermore, they are closely connected to other QAnon accounts but also maintain links to spreaders of fake news (\eg, thegatewaypundit.com), far-right activists (\eg, Tommy Robinson), and conservative thought leaders (\eg, Sean Hannity).

We also compare the in- and out-degree centrality of users in both groups. Here, in-degree centrality measures the number of times a user's post is reposted/commented by another user and vice versa. We find that QAnon supporters are more central compared to non-QAnon supporters. In particular, posts by QAnon supporters are reposted/commented by 29.78 other users on average while posts by non-QAnon supporters are only reposted/commented on by 16.66 other users. Similarly, QAnon supporters repost/comment on average 24.37 posts by other users compared to 17.01 for non-QAnon supporters. Overall, this underlines the strong impact of QAnon supporters on Parler network.

\noindent
\emph{What content do (non-)QAnon supporters share on Parler?}

\begin{figure}[ht]
	\captionsetup{position=top}
	\centering
	\includegraphics[width=0.8\linewidth]{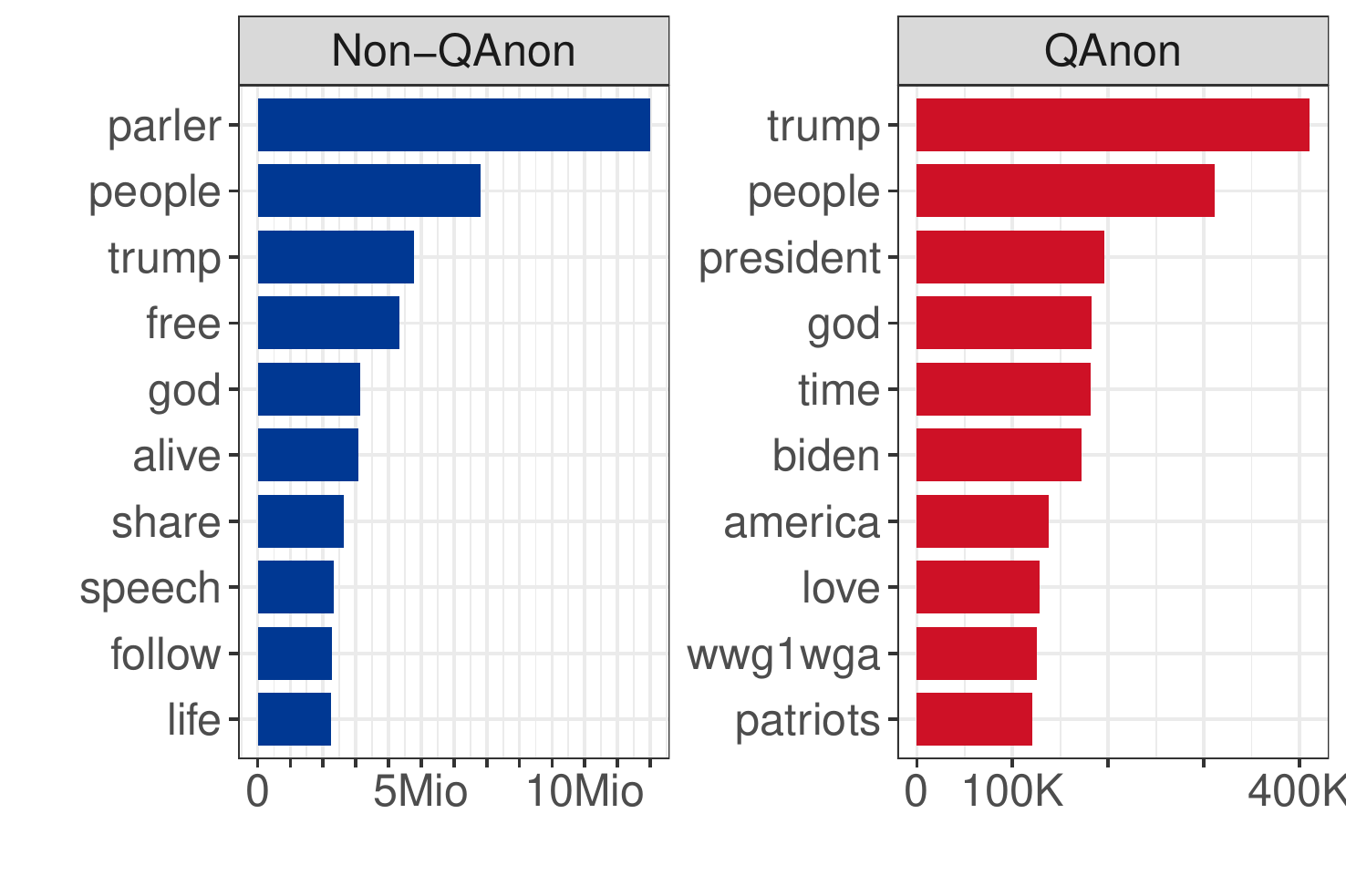}
	\caption{Top 10 most common words in Parler content (\ie, posts and comments) as a breakdown by QAnon and non-QAnon supporters.}
	\label{fig:frequent_words}
\end{figure}

We expect that QAnon supporters discuss QAnon-related topics in addition to other topics on Parler. To check this, we now compare the content (\ie, \emph{``what''} users write) shared by QAnon and non-QAnon supporters on Parler. In particular, we compare the most frequently used words by QAnon and non-QAnon supporters, which should point to how topics of interest vary across both groups. Fig.~\ref{fig:frequent_words} reports the 10 most frequent words appearing in content composed by QAnon vs. non-QAnon supporters. 

There are several similarities between QAnon and non-QAnon supporters. We find that both groups mention former U.S. President Donald Trump and ``god''. This may be expected: while many non-QAnon supporters do not self-disclose interest in the QAnon conspiracy theory, many of them are still conservatives (thus engaging in frequent discussions around politics or religion) \cite{Aliapoulios.2021}. Along these lines, we find that QAnon supporters frequently use words such as ``America'' and ``patriots'' which are often associated with a strong political polarization towards the right-wing. 

However, there are also differences. We find that QAnon supporters frequently share terms such as ``{WWG1WGA}'' (as a short form for ``Where we go one we go all'') that are inherently QAnon-specific. Evidently, QAnon supporters frequently engage in discussing conspiratorial content. 

Overall, the analysis shows that QAnon supporters discuss many conservative topics similar to those of regular users but, beyond that, also refer to words that are unique to the QAnon conspiracy theory. 

\subsection{Discriminating QAnon and non-QAnon Supporters Using Machine Learning (RQ3)}

To answer \textbf{RQ3}, we examine which of the different feature groups allows to discriminate QAnon and non-QAnon supporters using machine learning. We thus fit four XGBoost classifiers, each using one of the four different feature groups. In addition, we examine the predictive power of a classifier trained on a combination of all four feature groups.

Tbl.~\ref{tbl:results_xgboost} reports the prediction performance of the different classifiers. Mann–Whitney U-tests \cite{Mason.2002} are used to confirm that the performance is above that of a random guess. We find that all feature sets have predictive power based on which QAnon vs. non-QAnon supporters can be discerned. For each feature group, the respective ROC~AUC is above 0.50. Further, the improvement is statistically significant ($p<0.01$). Consistent with this, we find that the performance of a classifier trained based on a combination of all features achieves a ROC~AUC of 0.76. Again, this is statistically significant ($p<0.01$).

Beyond that, we observe several notable patterns. First, we find that user features have the highest discriminatory power wrt. distinguishing (non-)QAnon supporters (ROC~AUC = 0.74). This is followed by content features and linguistic features, where the performance of the classifier amounts to a ROC~AUC of 0.69 and 0.67, respectively. In contrast, network features have the lowest discriminatory power (ROC~AUC = 0.63).

We now check the feature importance for our model using all features. We find that user features (\eg, followers, account age) have overall large feature importance scores. In addition, several linguistic and content features (\eg, stance, SBERT 336) are also important. This corroborates our previous findings showing that QAnon supporters differ significantly from non-QAnon supporters along these dimensions (\eg, build larger friendship networks, use more hashtags).

\begin{table}
	\small
	\linespread{0.5}
	\begin{tabular}{p{0.28\linewidth}p{0.14\linewidth}p{0.14\linewidth}p{0.14\linewidth}p{0.05\linewidth}}
		\toprule
		\textbf{Input} & \textbf{ROC~AUC} & \textbf{Sensitivity} & \textbf{Specificity} & \textbf{F1} \\
		\midrule
		User features & 0.74$^{***}$ & 0.77 & 0.71 & 0.75 \\ 
		Linguistic features & 0.67$^{***}$ & 0.62 & 0.72 & 0.66 \\ 
		Network features & 0.63$^{***}$ & 0.62 & 0.65 & 0.63 \\
		Content features & 0.69$^{***}$ & 0.67 & 0.70 & 0.68 \\ 
		\midrule
		All features & 0.76$^{***}$ & 0.77 & 0.75 & 0.77 \\ 
		\bottomrule
		\multicolumn{5}{p{0.9\linewidth}}{\scriptsize{$p$-values are obtained using the Mann–Whitney U-test \cite{Mason.2002}: \scriptsize{$^{***}p<0.001$; $^{**}p<0.01$; $^{*}p<0.05$}}}\\
	\end{tabular}
	\caption{Performance of classifying users into QAnon (=1) vs. non-QAnon (=0) supporters based on different features. }
	\label{tbl:results_xgboost}
\end{table}

\textbf{Robustness Checks:} We check the robustness of our machine learning approach by conducting a series of additional analyses (details are in our GitHub): First, we check the predictive performance of each feature group with alternative classifiers. Overall, we find: XGBoost performs best. Various other classifiers reach a similar performance, and the ordering of the different feature groups remains the same. Second, we test the predictive performance with all possible combinations (\ie, powerset) of our four feature groups. In line with our main results, we find that combinations including user features exhibit higher ROC~AUC scores and that no combination can statistically significantly achieve higher predictive performance compared to a model based on the complete set of features. Third, we check if features with low predictive performance influence the overall classification performance and perform feature selection using a LASSO. Subsequently, we train a model on all features with non-zero coefficients. Our results indicate that the exclusion of features with low predictive power does not increase the classification performance (ROC~AUC=0.76). Nevertheless, this may yield a reduced model for applications in practice.

\section{Discussion}

\textbf{Relevance:} The QAnon conspiracy theory has been deemed as a significant threat to public security by the U.S. Federal Bureau of Investigation \cite{Amarasingam.2020}. Thus, a better understanding of QAnon supporters is necessary to identify potentially dangerous communities on the platform. To the best of our knowledge, this paper provides the first study profiling QAnon supporters on Parler.

\textbf{Summary of Findings:} Our findings contribute to the existing research on Parler by quantifying the number of QAnon supporters on the platform. Previously, it was frequently observed that the Parler social media platform hosts QAnon \cite[\eg,][]{Aliapoulios.2021, Baines.2021}; however, the actual size of the community has remained unclear. Here, our findings show that there is indeed a large community of self-reported QAnon supporters on Parler. Specifically, we find that around 5.5\,\% of the users in our sample are self-reported QAnon supporters. 

Our results further show that QAnon supporters differ from non-QAnon supporters across multiple dimensions. This is in line with social identity theory which predicts that the behavior of people differs according to their group membership \cite{Tajfel.1986}. For example, we find that QAnon supporters have a large impact on the platform as they are more active and maintain larger friendship and repost networks compared to non-QAnon supporters. As such, QAnon supporters appear to behave similarly to other abusive users on social media \cite{Ribeiro.2018}. Thereby, we establish a better understanding of the behavior of QAnon supporters in online communities. 

A potential reason for the different behavior of QAnon supporters might lie in the participatory nature of QAnon \cite{Zuckerman.2019} and the motivations behind conspiracy theory beliefs \cite{Sternisko.2020}. Psychological research shows that conspiracy theorists want to make sense of their environment \cite{Sternisko.2020}. In the case of QAnon, supporters collectively investigate the cabal and decipher Q drops. Such efforts require a high level of outreach and discussion to be successful \cite{Bar.2023b} and could thus lead to larger networks and activity on Parler. Along similar lines, the higher number of URLs shared by QAnon supporters might indicate an increased effort to explain their views. Furthermore, conspiracy theorists are often driven by social identity motives and adherents want to feel good about themselves / their group \cite{Sternisko.2020}. The frequent use of specific hashtags of QAnon supporters might indicate the development of a unique social identity, whereas the relatively more positive sentiment and the lower levels of toxicity, threat, and profanity might be related to a positive self-image of QAnon supporters. Of note, the differences for all these features are statistically significant, showing that QAnon supporters differ from the otherwise rather homogeneously conservative user base on Parler.

We further find that machine learning together with a representative set of features (chosen analogous to prior research \cite{Paul.2019, Rao.2010}) can discriminate QAnon vs. non-QAnon supporters with a ROC~AUC of up to 0.76. The performance is lower compared to studies profiling verified users \cite{Paul.2019} or detecting bots on Twitter \cite{Kudugunta.2018}. This implies that there is still some unexplained variance (beyond the information from user characteristics that is typically used for user profiling in social media). However, the lower ROC~AUC is also an indication of similarities among QAnon supporters and other users on Parler, who, for example, may often post somewhat similar content (\eg, posts with similar conservative viewpoints). Out of the different feature groups, we find that user features are especially discriminatory to distinguish QAnon vs. non-QAnon supporters, suggesting that both groups are characterized by different behavior (and not necessarily \emph{``how''} or \emph{``what''} users write). This is in line with other research profiling users on mainstream online communities such as Twitter, where user features consistently have large predictive power \cite{Kudugunta.2018, Paul.2019}.

\textbf{Implications:} The QAnon community is expanding globally \cite{Hoseini.2021} and poses a significant threat to public security \cite{Amarasingam.2020}. As such, the growing number of violent acts by QAnon supporters \cite{NationalConsortiumfortheStudyofTerrorismandResponsestoTerrorism.2021} that peaked in an attack on democracy with the storming of the U.S. Capitol on January~6,~2021 \cite{Hitkul.2021} call for action by research, platform owners, and policymakers. Here, our work provides first insights into the behavioral attributes of QAnon supporters. We demonstrate that machine learning in combination with a comprehensive set of features can help to identify QAnon supporters -- even on a social media platform that largely resembles a right-wing echo chamber. From a practical perspective, our features rely on data available for most social media platforms. As such, our machine learning framework is directly applicable to other platforms and may help surveillance and early detection of upcoming threats through a group of conspiracy theorists that have repeatedly been associated with violent incidents.

Moreover, it is concerning that QAnon supporters seem to use Parler to cultivate their social identity (\eg, by using different linguistic styles) while also growing larger friendship networks faster than other user groups. The latter may be expected as QAnon supporters chose the Parler platform for a particular reason (\eg, actively discussing politics based on a certain ideology rather than pure news consumption or curiosity). However, it also renders it likely that false information spreads particularly fast and viral among QAnon supporters due to their larger reach and central role in the network. Also, we see that community mechanisms to control information (\ie, upvotes and downvotes) may not be functioning as desired. Upvotes are distributed fairly similarly for both QAnon supporters and other users on Parler. However, more importantly, comments from QAnon supporters receive \emph{considerably} fewer downvotes. This may exacerbate and even reinforce the spread of false information due to the absence of content moderation (as well as having a segregated ``echo chamber'' platform with users originating primarily from a single, right-leaning ideology).

\textbf{Limitations and Future Research:} As with other research, ours is not free of limitations that offer opportunities for future research: (1)~We identified QAnon supporters based on an extensive series of keywords consistent with earlier research \cite{Sharma.2022}. Previously, the reliability of such an approach was unclear (\eg, users might express their opposition to QAnon and still get classified as supporters). However, our validation study confirms that the approach is highly accurate. (2)~We infer QAnon supporters based on their user bios, while future research could identify QAnon supporters based on posts and comments, though this might also include users arguing against QAnon and may thus provide less reliable labels. (3)~Our analysis is based on Parler, which has attracted a large community of QAnon supporters and which makes it particularly relevant for research to understand differences between QAnon vs. non-QAnon supporters. However, QAnon supporters might behave differently on mainstream social media. As a result, taking a cross-platform perspective presents a promising avenue for future research. (4)~Our data provides a static snapshot of Parler. Hence, future research should study how the characteristics of QAnon supporters change over time. (5)~We compare QAnon vs. non-QAnon supporters. Yet, there may also be further heterogeneity among QAnon supporters, which could be analyzed by future research.

\section{Conclusion}

The social media platform Parler has emerged into a prominent fringe community where a significant proportion of the user base are self-reported supporters of QAnon. Yet, little is known about QAnon supporters on Parler. To fill this void, we analyze a large-scale public snapshot of Parler, based on which we profile QAnon supporters. Self-reported QAnon supporters make up a significant portion (5.5\,\%) of the user base on Parler. These users are significantly different from non-QAnon supporters on Parler. Following social identity theory, the self-reported appraisal of QAnon manifests in different online behavior such as larger friendship networks and greater activity. These differences allow machine learning to discriminate QAnon vs. non-QAnon supporters. Here, user features such as the size of the friendship network or activity are more discriminatory compared to \emph{how} or \emph{what} users write. Our machine learning framework may thus allow for real-time surveillance and early warnings.

\section{Ethics Statement}

This research did not involve interventions with human subjects, and, thus, no approval from the Institutional Review Board was required by the author institutions. All analyses are based on publicly available data and we do not make any attempt to track users across different platforms. We neither de-anonymize nor de-identify their accounts. Furthermore, all analyses conform with national laws. To respect privacy, we explicitly do not publish usernames in our paper (except for celebrity profiles) and only report aggregate results.

\bibliography{literature}

\end{document}